\documentclass{aa}
\usepackage{graphicx}
\usepackage{color}

\usepackage{subfig}
\usepackage{float}

\begin{document}
   \title{Magnetic field structure in single late-type giants: The weak G-band giant 37 Comae from 2008 to 2011\thanks{Based on observations obtained at the Bernard Lyot T\'elescope (TBL, Pic du Midi, France) of the Midi-Pyr\'en\'ees Observatory which is operated by the Institut National des Sciences de l'Univers of the Centre National de la Recherche Scientifique of France and Universit\'e de Toulouse, and at the Canada-France-Hawaii Telescope (CFHT) which is operated by the National Research Council of Canada, the Institut National des Sciences de l'Univers of the Centre National de la Recherche Scientifique of France, and the University of Hawaii.}}


   \author{S. Tsvetkova\inst{1}, P. Petit\inst{2,3}, R. Konstantinova-Antova\inst{1,3}, M. Auri\`ere\inst{2,3}, G.A. Wade\inst{4}, A. Palacios\inst{5}, C. Charbonnel\inst{6,2}, N.A. Drake\inst{7,8}}

   \offprints{stsvetkova@astro.bas.bg}

   \institute{Institute of Astronomy and NAO, Bulgarian Academy of Sciences, 72 Tsarigradsko shose, 1784 Sofia, Bulgaria\\
              \email{stsvetkova@astro.bas.bg}
     \and CNRS, UMR 5277, Institut de Recherche en Astrophysique et Plan\'etologie, 14 Avenue Edouard Belin, 31400 Toulouse, France
     \and Universit\'e de Toulouse, UPS-OMP, Institut de Recherche en Astrophysique et Plan\'etologie, Toulouse, France
     \and Department of Physics, Royal Military College of Canada, PO Box 17000, Station `Forces', Kingston, Ontario, Canada K7K 4B4
     \and LUPM, University of Montpellier, CNRS, Place Eug\`ene Bataillon - cc072, 34095 Montpellier, France
     \and Department of Astronomy, University of Geneva, Chemin des Maillettes 51, 1290 Versoix, Switzerland
     \and Saint Petersburg State University, Universitetski pr. 28, Saint Petersburg 198504, Russia
     \and Observat\'orio Nacional/MCTI, Rua General Jos\'e Cristino 77, 20921-400, Rio de Janeiro, Brazil\\
              }

   \date{Received ; }


  \abstract
   {}
   {This work studies the magnetic activity of the late-type giant 37 Com. This star belongs to the group of weak G-band stars that present very strong carbon deficiency in their photospheres. The paper is a part of a global investigation into the properties and origin of magnetic fields in cool giants.}
   {We use spectropolarimetric data, which allows the simultaneous measurement of the longitudinal magnetic field $B_{l}$, line activity indicators (H$\alpha$, Ca\,{\sc ii} IRT, S-index) and radial velocity of the star, and consequently perform a direct comparison of their time variability. Mean Stokes $V$ profiles are extracted using the least squares deconvolution (LSD) method. One map of the surface magnetic field of the star is reconstructed via the Zeeman Doppler imaging (ZDI) inversion technique.}
   {A periodogram analysis is performed on our dataset and it reveals a rotation period of 111 days. We interpret this period to be the rotation period of 37 Com. The reconstructed magnetic map reveals that the structure of the surface magnetic field is complex and features a significant toroidal component. The time variability of the line activity indicators, radial velocity and magnetic field $B_{l}$ indicates a possible evolution of the surface magnetic structures in the period from 2008 to 2011. For completeness of our study, we use customized stellar evolutionary models suited to a weak G-band star. Synthetic spectra are also calculated to confirm the peculiar abundance of 37 Com.}
   {We deduce that 37 Com is a 6.5~$M_{\odot}$ weak G-band star located in the Hertzsprung gap, whose magnetic activity is probably due to dynamo action.}

   \authorrunning{Tsvetkova et al.}
   \titlerunning{Magnetic field structure of 37 Comae}
   \maketitle
%

\section{Introduction}

The detailed study of magnetic fields in single late-type giants via Zeeman signatures began just a few years ago (except for the case of FK Com type stars, Petit et al. 2004). First detections of magnetic fields for rapidly rotating giants were reported by Konstantinova-Antova et al. (2008~a,~b, 2009) and for some slowly rotating giants by Auri\`ere et al. (2009~a,~b). Longitudinal magnetic fields of the order of several gauss to several tens of gauss were measured at the surfaces of these giants by employing the least squares deconvolution (LSD) method (Donati et al. 1997).

A sample of single late-type giants, which are reported in the literature to present evidence for activity, were studied by Auri\`ere et al. (2015). The authors detected surface magnetic fields in almost all of the giants from that sample. Most of the detected stars are situated at the first dredge-up phase and core helium-burning phase. The results from the study indicate that the magnetic activity of single late-type giants could be due to the operation of an $\alpha-\omega$ dynamo with Rossby numbers (determined semi-empirically) that are smaller than unity. In addition, the authors reported about four stars that do not fit to the general trends (for example the rotation - field strength correlation), and which are suspected of being descendants of Ap stars (chemically peculiar magnetic stars of spectral types A and late B on the main sequence).

A second survey has just been completed. It involves the spectropolarimetric study of 45 G, K, and M giants with $V \leq 4$~mag in the Solar vicinity. Forty-one of 45 stars were observed and 44 \% of them are Zeeman detected. The results are reported by Konstantinova-Antova et al. (2014) and Konstantinova-Antova et al. (2017 in prep.). Poor weather resulted in no observations being acquired for the four objects.

There are only a few single late-type giants, for which sufficient observations with sufficiently good phase coverage have been collected to map their surface magnetic fields via the Zeeman Doppler imaging (ZDI) technique. These giants are V390 Aur (at the first dredge-up phase, Konstantinova-Antova et al. 2012), EK Eri (at the first dredge-up phase, Auri\`ere et al. 2011), $\beta$ Ceti (at the core He-burning phase, Tsvetkova et al. 2013), and OU And and 31 Com (both at the Hertzsprung gap, Borisova et al. 2016). The magnetic activity of V390 Aur is very likely due to the operation of a dynamo and its ZDI map reveals a complex surface magnetic topology. On the contrary, the magnetic activity of EK Eri and $\beta$ Ceti is more likely due to fossil fields interacting with convection. Both giants have magnetic fields with simples topologies, combined with slow rotation. They are suspected of being descendants of magnetic Ap stars. The magnetic field of 31 Com has a dynamo origin, while OU And is interpreted as being an Ap star descendant with remnant fast rotation, a magnetic fossil field, and a developing dynamo component (Borisova et al. 2016).


In this way, EK Eri and $\beta$ Ceti, and in addition 14 Ceti (entering the Hertzsprung gap; Auri\`ere et al. 2012), represent illustrations that magnetic fields in Ap stars may survive at more advanced evolutionary stages after the main sequence.

The present paper is focused on the late-type giant 37 Com (HD 112989). It is classified as a single G9~III giant by Keenan \& McNeil (1989) and later confirmed by de Medeiros et al. (1999). Earlier studies indicate it is an intermediate-mass star with a mass around 5~$M_{\odot}$ (de Medeiros et al. 1999, Tokovinin 2008, Adamczak \& Lambert 2013, Auri\`ere et al. 2015). The star shows significant photometric and Ca\,{\sc ii} K\&H emission core variability (de Medeiros et al. 1999), which is interpreted as indicative of magnetic activity. The first magnetic field detection for 37 Com was reported by Konstantinova-Antova et al. (2009). It is moreover listed as a weak G-band giant (hereafter wGb) by Lambert \& Ress (1981) and by Adamczak \& Lambert (2013), meaning that it exhibits a very strong underabundance of carbon that is evidenced in its spectrum by a very weak CH G-band at 4300 \AA.

Recently, Tokovinin (2008) indicated in his catalog that 37 Com is the primary star in a wide triple system. However, the synchronization effect appears to play no role for its fast rotation ($v \sin i = 11 \pm 1$~km\,s$^{-1}$, according to the catalog of de Medeiros \& Mayor 1999) and activity. The case of the giant V390 Aur is similar: it is also the primary of a wide triple system, but it has been shown that it may be considered as an effectively single giant (Konstantinova-Antova et al. 2008~b), so the other components of the system did not affect the magnetic study and the process of reconstructing the magnetic map of the star (Konstantinova-Antova et al. 2012).

Our study of the magnetic field and activity of 37 Com includes high-resolution spectropolarimetric data that were collected in the period April 2008 -- February 2011. In Sect. 2 we describe the observations and data reduction. Section 3 reports our results -- the first Zeeman Doppler imaging map of the star, time variability of the longitudinal magnetic field, line activity indicators, and radial velocity. In Sect. 4, synthetic spectra are employed to derive the atmospheric parameters, metallicity, the abundances of light elements and Li, and the carbon isotopic ratio $^{12}$C/$^{13}$C. In Sect. 5, we determine the mass, evolutionary status, initial rotation velocity and Rossby number of the star. Finally, Sect. 6 contains our interpretation of the results, and our conclusions are given in Sect. 7.

\section{Observations and data reduction}

Observational data were obtained with two twin fiber-fed \'echelle spectropolarimeters -- Narval (Auri\`ere 2003), which operates at the 2-m T\'elescope Bernard Lyot (TBL) at Pic du Midi Observatory, France, and ESPaDOnS (Donati et al. 2006~a), which operates at the 3.6-m Canada-France-Hawaii Telescope (CFHT). In polarimetric mode, both spectropolarimeters have a spectral resolving power of about 65\,000 and a nearly continuous spectrum coverage from the near-ultraviolet (at about 370~nm) to the near-infrared domain (at about 1050~nm) in a single exposure, with 40 orders aligned on the CCD frame by 2 cross-disperser prisms. The Stokes $I$ (unpolarized light) and Stokes $V$ (circular polarization) parameters are simultaneously measured using a sequence of four sub-exposures, between which the retarders (Fresnel rhombs) are rotated in order to exchange the beams in the instrument and to reduce spurious polarization signatures (Semel et al. 1993).

The automatic reduction software LibreEsprit (Donati et al. 1997), which includes the optimal extraction of the spectrum, wavelength calibration, correction to the heliocentric frame, and continuum normalization, is applied to all observations. We then use the LSD multiline technique (Donati et al. 1997), which enables the averaging of about 14\,000 spectral lines in the case of 37 Com, to detect weak polarized Zeeman signatures. In this way, mean Stokes $I$ and $V$ line profiles are generated. The line mask is created from the VALD atomic data base (Kupka et al. 1999) and is calculated for an effective temperature $T_{\rm eff} = 4500$~K, $\log g = 2.5$ and a microturbulence of 2~km\,s$^{-1}$, consistent with the literature data for 37 Com (Brown et al. 1989, McWilliam 1990, Soubiran et al. 2010, Adamczak \& Lambert 2013).

\begin{table*}[!htc]
\centering
\begin{center}

\caption{Journal of observations and measurements of the line activity indicators, the longitudinal magnetic field $B_{l}$ and radial velocity for 37 Com.}

\label{table:1}

\centering
\begin{tabular}{c c c c c c c c c c c}
\hline\hline

     &Date&    HJD      & Rot.& S/N &         &                &       &$B_{l}$&$\sigma$&    RV        \\
Inst.&UT  &2\,450\,000 +&phase&(LSD)&H$\alpha$&Ca\,{\sc ii} IRT&S-index&  [G]  &  [G]   &[km\,s$^{-1}$]\\
\hline
(1)  & (2)&  (3)        & (4) & (5) & (6)     &     (7)        & (8)   &  (9)  &  (10)  & (11)\\
\hline
\\
N & 04 Apr 08 & 4560.521 & 0.000 & 63~139 & 0.254 & 0.164 & 0.343 &  4.6 & 0.7 & -14.41\\
N & 07 Apr 08 & 4563.552 & 0.027 & 45~476 & 0.254 & 0.165 & 0.346 &  5.2 & 0.9 & -14.41\\
N & 25 Feb 09 & 4887.587 & 2.947 & 60~705 & 0.292 & 0.181 & 0.394 & -1.8 & 0.7 & -14.46\\
E & 27 Jan 10 & 5224.053 & 5.978 & 63~917 & 0.267 & 0.175 & 0.371 & -2.0 & 0.7 & -14.56\\
E & 27 Jan 10 & 5224.079 & 5.978 & 69~999 & 0.267 & 0.175 & 0.372 & -0.9 & 0.6 & -14.56\\
E & 29 Jan 10 & 5226.153 & 5.997 & 74~618 & 0.264 & 0.173 & 0.383 & -0.7 & 0.6 & -14.55\\
E & 29 Jan 10 & 5226.167 & 5.997 & 80~645 & 0.264 & 0.174 & 0.375 & -1.9 & 0.5 & -14.55\\
E & 02 Feb 10 & 5230.143 & 6.033 & 69~626 & 0.265 & 0.170 & 0.369 & -2.5 & 0.6 & -14.63\\
N & 13 Mar 10 & 5269.492 & 6.387 & 44~481 & 0.299 & 0.192 & 0.397 &  5.2 & 0.9 & -14.61\\
N & 23 Mar 10 & 5278.528 & 6.468 & 52~064 & 0.317 & 0.185 & 0.422 &  4.9 & 0.8 & -14.52\\
N & 05 Apr 10 & 5292.502 & 6.594 & 32~658 & 0.293 & 0.181 & 0.377 &  4.6 & 1.3 & -14.58\\
N & 11 Apr 10 & 5297.537 & 6.640 & 52~807 & 0.281 & 0.175 & 0.365 &  4.0 & 0.8 & -14.55\\
N & 16 Apr 10 & 5303.483 & 6.693 & 53~299 & 0.291 & 0.180 & 0.390 &  3.7 & 0.8 & -14.53\\
N & 24 Apr 10 & 5310.577 & 6.757 & 56~390 & 0.330 & 0.194 & 0.433 &  3.3 & 0.7 & -14.42\\
N & 21 Jun 10 & 5369.374 & 7.287 & 46~761 & 0.280 & 0.186 & 0.368 &  6.5 & 0.9 & -14.44\\
E & 21 Jul 10 & 5398.740 & 7.551 & 58~530 & 0.265 & 0.166 & 0.385 &  4.9 & 0.7 & -14.58\\
N & 10 Dec 10 & 5540.753 & 8.831 & 44~849 & 0.300 & 0.187 & 0.384 &  3.3 & 1.0 & -14.51\\
E & 15 Dec 10 & 5546.145 & 8.880 & 79~635 & 0.313 & 0.185 & 0.407 &  1.7 & 0.5 & -14.38\\
E & 28 Dec 10 & 5559.171 & 8.997 & 53~027 & 0.306 & 0.191 & 0.417 & -3.6 & 0.8 & -14.37\\
E & 30 Dec 10 & 5561.160 & 9.015 & 77~358 & 0.305 & 0.189 & 0.419 & -1.8 & 0.5 & -14.36\\
N & 04 Jan 11 & 5565.682 & 9.056 & 51~706 & 0.293 & 0.195 & 0.397 & -1.9 & 0.8 & -14.48\\
N & 14 Jan 11 & 5575.669 & 9.146 & 26~252 & 0.284 & 0.191 & 0.379 & -2.1 & 1.6 & -14.53\\
N & 23 Jan 11 & 5584.653 & 9.226 & 52~680 & 0.272 & 0.179 & 0.369 &  2.1 & 0.8 & -14.50\\
N & 01 Feb 11 & 5593.645 & 9.307 & 36~492 & 0.284 & 0.190 & 0.376 &  5.4 & 1.1 & -14.49\\
\\
\hline
\hline
\end{tabular}
\end{center}
\end{table*}

We obtained 24 spectra (on 22 different nights) of 37 Com, which cover the time interval from April 2008 to February 2011. The journal of observations and the measurements of the line activity indicators (H$\alpha$, the 854.2~nm line of the Ca\,{\sc ii} IRT, and S-index), longitudinal magnetic field $B_{l}$ and radial velocity (RV) are presented in Table~\ref{table:1}. In the first column of Table~\ref{table:1}, the capital letters N and E stand for Narval and ESPaDOnS, respectively. The fourth column presents the rotational phase of each observation, calculated assuming a rotation period of the giant of 111 days (the period search analysis is described in Sect.~\ref{sec:periodsearch}). The fifth column gives the signal-to-noise ratio (S/N) of each Stokes $V$ LSD profile (calculated according to the following values of the normalization parameters -- normalized line depth equal to 0.59, wavelength equal to 627~nm and Land\'e factor equal to 1.20).

The line-of-sight-component of the stellar magnetic field integrated over the visible stellar disc, $B_{l}$, was computed using the first-order moment method (Rees \& Semel 1979, Donati et al. 1997, Wade et al. 2000~a, b). The first moment was computed between radial velocity boundaries set to encompass the whole velocity span of Stokes $V$ signatures. For 37 Com that integration window has velocity boundaries of $\pm$ 20~km\,s$^{-1}$ around the line center of LSD profiles. The longitudinal magnetic field $B_{l}$ (expressed in gauss) was computed from both LSD Stokes $I$ and $V$ profiles with the following equation:

\begin{equation}
 B_{l} =-2.14\times10^{11}\frac{\int vV(v)\,dv}{\lambda gc \int [1-I(v)]\,dv},
\end{equation}
where $v$ (km\,s$^{-1}$) is the radial velocity in the stellar restframe, $\lambda$ (in nm) is the normalization wavelength (627~nm for 37 Com), $g$ is the Land\'e factor (here 1.20), and c (km\,s$^{-1}$) is the light velocity in vacuum. The measured values of $B_{l}$ and their uncertainties for 37 Com are given in Table~\ref{table:1}.

An additional indicator of magnetic activity, which gives information regarding chromospheric variations, is the S-index, defined from the Mount Wilson Survey (Duncan et al. 1991). It is a dimensionless ratio between the emission in the line cores of Ca\,{\sc ii} K\&H to the emission in two nearby quasi-continuum bandpasses, which are close to the H and K lines. The procedure of measuring the S-index is described in detail by Marsden et al. (2014), who use calibration coefficients adjusted for solar-type stars. Auri\`ere et al. (2015) follow the same procedure, but with coefficients adjusted particularly for cool giants, according the following equation:

\begin{equation}
 S_{index} = \frac{aF_{H} + bF_{K}}{cF_{R_{HK}} + dF_{V_{HK}}} + e,
\end{equation}
where $F_{H}$ and $F_{K}$ are the fluxes in triangular bandpasses with a full width at half maximum (FWHM) of 0.1~nm centered on the Ca\,{\sc ii} H\&K line cores, and $F_{R_{HK}}$ and $F_{V_{HK}}$ are the fluxes in two 2~nm-wide rectangular bandpasses centered on 400.107 and 390.107~nm, respectively, which are used for the continuum flux at the red and blue sides of the H\&K lines. The coefficients are calculated independently for Narval and ESPaDOnS. To measure the S-index of 37 Com, we use the coefficients given by Auri\`ere et al. (2015) for cool giants and we list them in Table~\ref{table:coefficients}. Statistical error bars on the S-index are not provided here since they turn out to be dominated by inaccuracies in the default continuum normalization performed by LibreEsprit of the spectral orders containing the Ca H and K spectral lines.

\begin{table}[!hc]
\centering
\caption{Calibration coefficients of S-index for cool giants, given by Auri\`ere et al. (2015).}
\label{table:coefficients}
\begin{tabular}{c c c}
\hline
Coefficients & Narval & ESPaDOnS \\
\hline
a & -1.55               & 0.77                \\
b & $1.37\times10^1$    & 7.70                \\
c & 3.57                & 5.15                \\
d & 5.02                & 1.09                \\
e & $2.21\times10^{-2}$ & $-1.9\times10^{-2}$ \\
\hline
\end{tabular}
\end{table}

The spectral coverage of the two instruments (370 -- 1050~nm) also enables us to measure two more line activity indicators, which are the absorption lines H$\alpha$ (656.3~nm) and Ca\,{\sc ii} IRT (854.2~nm line). The intensities of both lines are measured relative to the continuum. The statistical error bars for the H$\alpha$ line are in the range from 0.005 to 0.010 and, for the line Ca\,{\sc ii} IRT, are in the range from 0.006 to 0.011.

We also measured the radial velocity (RV) of the LSD Stokes $I$ line profiles (Table~\ref{table:1}) by performing the $\chi^2$ adjustment of a Gaussian function on the line core and then taking the central velocity of the Gaussian as our RV value. Statistical error bars on RV measurements are not considered here, since the actual uncertainty on this quantity is mostly dominated by the biases induced by the final wavelength calibration provided by LibreEsprit, based on the position of telluric lines in the spectrum. This intrinsic accuracy is estimated to be of the order of 30~m\,s$^{-1}$ (Moutou et al. 2007).

\section{Results}
\label{sec:magneticstudy}

\subsection{ZDI -- the general model}

The magnetic topology of 37 Com was reconstructed by the Zeeman Doppler Imaging tomographic method (ZDI; Semel 1989, Donati \& Brown 1997, Donati et al. 2006~b), using the rotational modulation of Stokes $V$ signatures. The method considers that the surface vectorial magnetic field is projected onto a spherical harmonics frame, which easily enables us to distinguish between the poloidal and toroidal components of the surface magnetic field (Donati et al. 2006~b). This method performs iterative fitting to the observed time series of polarized LSD profiles by a simulated set of Stokes $V$ profiles that are computed for an identical sequence of rotational phases. The synthetic Stokes profiles are calculated from an artificial star, whose surface is divided into a grid of 2\,000 rectangular pixels of roughly similar area. Each surface pixel is associated with a local Stokes $I$ and $V$ profile. The maximum entropy reconstruction code is from Brown et al. (1991), who implement the algorithm for maximum entropy optimization from Skilling \& Bryan (1984).

The local synthetic Stokes $I$ line profile is assumed to possess a Gaussian shape, with a depth and width adjusted to achieve the best fit between synthetic and observed line profiles. This resulted in a width of $10^{-2}$~nm and a depth of 0.295 (in units of the continuum level). The Land\'e factor and wavelength of the local profile are kept equal to the normalization factors used to compute the LSD pseudo-profiles. For a given magnetic field strength and orientation for each pixel, local Stokes $V$ profiles are calculated under the weak-field assumption (Donati et al. 2003). The linear limb darkening coefficient is set to 0.75, in agreement with Claret \& Bloemen (2011).

Several input parameters are needed to reconstruct the magnetic topology of 37 Com: the rotation period, the inclination angle, and the differential rotation parameters, which are all discussed in the next sections. The adopted value for the projected rotational velocity of 37 Com is $v \sin i = 11 \pm 1$~km\,s$^{-1}$, according to the catalog of de Medeiros \& Mayor (1999). The spherical harmonics expansion is limited to $l \le 10$, since no improvement is found for the fits between modeled and observed LSD profiles whenever higher values of $l$ are considered in the model.

\subsection{The rotation period of 37 Com}
\label{sec:periodsearch}

First, we ran the period search over observational data from January 2010 to February 2011. About 300 ZDI models were computed, assuming a different rotation period of the star for each model, with a uniform sampling of rotation periods in the 30- to 200-day interval. Forcing a constant information content (i.e. the same average magnetic field strength) in all 300 computed magnetic models, we compared them to the observed Stokes $V$ profiles and analyzed the reduced $\chi^2$ ($\chi_{r}^2$ hereafter) as a function of period. Our best magnetic model, identified by the lowest $\chi_{r}^2$ value, suggests a possible rotation period of 111 days, which we use to reconstruct the magnetic map of the giant 37 Com. This approach to search for a period was proposed by Petit et al. (2002).

\begin{figure}[!htc]
    \centering
    \includegraphics[width=10cm]{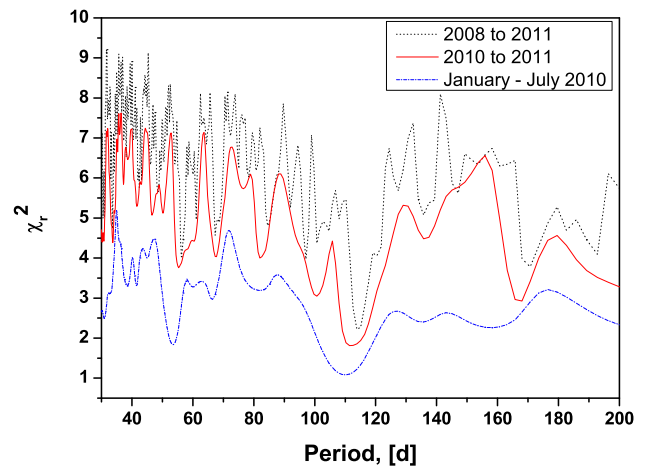}
     \caption{Periodograms of 37 Com, obtained for three different lengths of the datasets.}
   \label{fig:periodogram}
   \end{figure}

To check the value of 111 days, we again ran the period search for two different lengths of the dataset, searching in the same interval of 30-200 days. The first contains all of our observational data (April 2008 to February 2011) -- only three additional points are included (two points from 2008 and one point from 2009) relative to the previous dataset. The lowest $\chi_{r}^2$ value is achieved for a rotation period of 114 days. The second contains the data from January 2010 to July 2010, which excludes the gap of more than four months without observations from August -- December 2010, and for which we achieved a very good distribution of observations. The lowest $\chi_{r}^2$ value this time is achieved for a rotation period of 110 days. All the periodograms are shown in Fig.~\ref{fig:periodogram}.

A photometric period of 67 days for 37 Com is reported by Strassmeier \& Hall (1988), but with very low significance (according to the authors). Our results from the periodogram analysis do not confirm their rotation period, as can be seen in Fig.~\ref{fig:periodogram}. Besides the periodograms, we reconstructed one ZDI map with a period of 67 days. The fit between observed and modeled Stokes $V$ profiles was very poor, rejecting the value of 67 days as a possible rotation period of the giant 37 Com.

Adopting a rotation period of 111 days, we calculated the inclination angle $i = 39^\circ$, using the values of $v \sin i = 11$~km\,s$^{-1}$ (de Medeiros \& Mayor 1999) and a radius of the star of $38.6~R_{\odot}$. The values of the stellar radius and mass are discussed in Sect.~\ref{sec:evolution}. The uncertainty of $5^\circ$ for the inclination angle was derived based on the uncertainties in stellar radius, $v \sin i$ and period.

All data are therefore phased according to the following ephemeris with a rotation period of 111 days:

\begin{equation}
 HJD=2454560.52140 + 111\, \phi,
\end{equation}
where HJD is the heliocentric Julian date of the observations and $\phi$ is the rotational cycle. Calculated in this way, rotational phases are listed in Table~\ref{table:1} (column 4).



\subsection{The surface differential rotation of 37 Com}
\label{sec:diffrotparameters}

We use the maximum-entropy image reconstruction method to search for surface differential rotation from circularly polarized data (Petit et al. 2002). We assume here a solar-like differential rotation law, expressed as

\begin{equation}
\label{eq:diffrot}
 \Omega (l) = \Omega_{eq} - d \Omega \, \sin^{2}l,
\end{equation}
where $\Omega (l)$ is the rotation rate at latitude $l$, $\Omega_{\rm eq}$ is the rotation rate of the equator, and $d \Omega$ is the difference in the rotation rate between the pole and equator.

The basic idea of the method is that a two-dimensional (2D) parameter space exists, which consists of the pairs ($\Omega_{\rm eq}, d \Omega$). For each of the pairs, there is one reconstructed magnetic image of the stellar surface. The optimal differential rotation parameters minimize the information content (mean magnetic field intensity over the stellar surface) of the reconstructed image. To select the values for the optimal pair, a 2D paraboloid fit to the $\chi_{r}^2$ vs. both $\Omega_{\rm eq}$ and $d \Omega$ is used.

\begin{figure}[!hc]
      \centering
      \includegraphics[width=8cm]{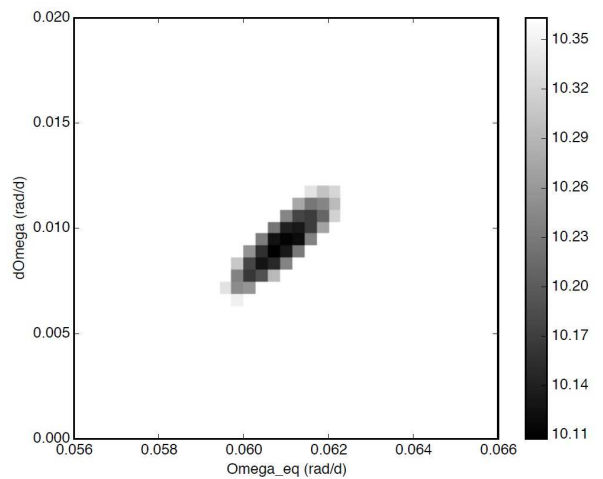}
      \caption{Variation of the average magnetic field of ZDI models (expressed in gauss) across the $\Omega_{\rm eq} - d \Omega$ plane (here from a series of $30 \times 30$ models). According to the maximum entropy principle, the more likely set of differential rotation parameters is the one that minimizes the average magnetic field of the ZDI map.}
   \label{fig:diffrot}
   \end{figure}

The error bars for the differential rotation parameters are computed by reconstructing different entropy landscapes similar to the one of Fig.~\ref{fig:diffrot}, but with ZDI main input parameters ($v \sin i$, inclination) varied over their typical uncertainties. The scatter in the position of our best model in the $\Omega_{\rm eq} - d \Omega$ plane is taken as a proxy of the determination uncertainty of the differential rotation parameters.

Using a square grid of 900 ZDI models, the best fit to our data (from January 2010 to February 2011) for 37 Com is achieved for the values of the parameters $\Omega_{\rm eq} = 0.061 \pm 0.001$~rad/d and $d \Omega = 0.009 \pm 0.001$~rad/d. The resulting average magnetic landscape is shown in Fig.~\ref{fig:diffrot}.

These values of the differential rotation parameters are used to reconstruct the magnetic map of 37 Com.

\subsection{The magnetic map and the model for 37 Com}
\label{sec:ZDImap}

\begin{figure}[!hc]
    \centering
    \includegraphics[width=6cm]{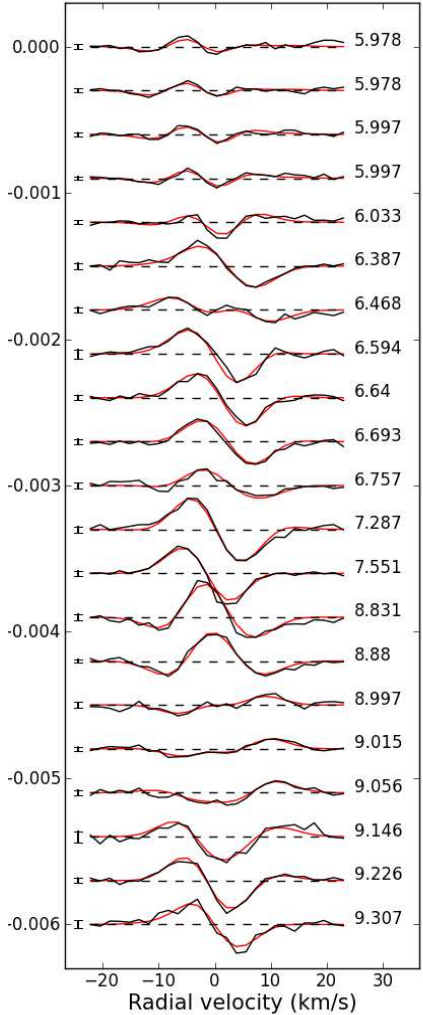}
     \caption{Normalized Stokes $V$ profiles of 37 Com in the period from January 2010 to February 2011 -- observed profiles (black lines); synthetic profiles (red lines); zero level (dashed lines). All profiles are shifted vertically for display purposes. The rotational phases of observations are indicated in the right part of the plot and the error bars are on the left of each profile.}
   \label{fig:37ComStokesV}
   \end{figure}

\begin{figure*}
    \centering
    \includegraphics[width=12cm]{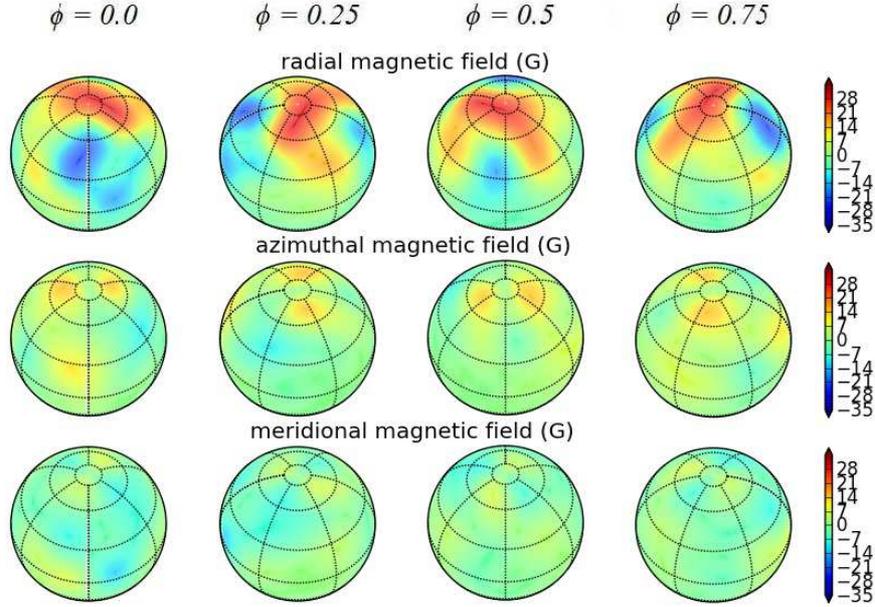}
     \caption{The magnetic map of 37 Com in the period from January 2010 to February 2011, derived from Stokes $I$ and $V$ profiles. The three rows illustrate the field components in spherical coordinates (from top to bottom -- radial, azimuthal, meridional). The four columns correspond to four rotational phases, which are indicated at the top of the figure. The magnetic field strength is expressed in gauss.}
   \label{fig:37ComZDImap}
   \end{figure*}

In Fig.~\ref{fig:37ComStokesV}, we show 21 Stokes $V$ profiles (January 2010 -- February 2011), which are used to reconstruct the final magnetic map. Their rotational phases are indicated beside each profile and are the same as in column 4, Table~\ref{table:1}. Starting from the top, Stokes $V$ profiles 1 to 5 (January 27 to February 2, 2010) and those from 16 to 19 (December 28, 2010 to January 14, 2011) correspond to negative values of $B_{l}$ (Table~\ref{table:1}, column 9). These line profiles have rotational phases around $\phi=0$, where a spot of radial field with negative polarity dominates, as shown by the magnetic map in Fig.~\ref{fig:37ComZDImap}.

The seventh Stokes $V$ profile (March 23, 2010; rotational phase 6.468) has a more complex shape than the others. This is due to the fact that despite the positive sign of $B_{l}$, a small spot with negative polarity is visible and contributes to the global profile, which is otherwise the mainly the result of the positive polar spot.

The other Stokes $V$ profiles in Fig.~\ref{fig:37ComStokesV} have simple shapes with positive blue lobes and negative red lobes, which is in good agreement with the measured field of positive polarity.

The three spectra from 2008 and 2009 were not used to reconstruct the final ZDI map of 37 Com in Fig.~\ref{fig:37ComZDImap}. The values of $B_{l}$ for 2008 have positive signs and their rotational phases are around $\phi=0$, which does not follow the general tendency for intervals with positive and negative $B_{l}$. The rotational phase of the data point from 2009 is close to $\phi=0$, but this time the value of $B_{l}$ has a negative sign, following the general tendency for the signs of $B_{l}$. In fact we ran the ZDI code adding these three spectra from 2008 and 2009 to the others using the same parameters as those used to make the map in Fig.~\ref{fig:37ComZDImap}. The fit between the observed and modeled profiles from 2008 and 2009 became worse, especially for the profile from 2009. Of course, this affected the $\chi_{r}^2$, resulting in a higher value (as can be seen directly in the higher $\chi_{r}^2$ values of the periodogram of Fig.~\ref{fig:periodogram} that were obtained using these data).

Probably, this result is indicative of an evolution of the large-scale surface magnetic structures of 37 Com, in a way that cannot be modeled through solar-like differential rotation alone. In support of this impression, we note two profiles that have the same rotation phase (January 29, 2010 with a phase 5.997 and December 28, 2010 with a phase 8.997), but exhibiting small differences in their shapes. It appears that these changes, in this specific case at least, are convincingly modeled by our differentially-rotating artificial star, so that we are able to combine our data from several rotational cycles to achieve a better phase coverage.

The ZDI map of 37 Com is shown in Fig.~\ref{fig:37ComZDImap}. This is reconstructed with an inclination angle $i = 39^\circ$ (Sect.~\ref{sec:periodsearch}) and differential rotation parameters $\Omega_{\rm eq} = 0.061$~rad/d and $d \Omega = 0.009$~rad/d (Sect.~\ref{sec:diffrotparameters}).

The ZDI map has a $\chi_{r}^2 = 1.0$. The magnetic geometry is dominated by the poloidal component of the magnetic field, which contains about 88~\% of the reconstructed magnetic energy. The amount of the poloidal magnetic energy, which is stored in the dipolar ($l = 1$) component, is about 22~\% and in the quadrupolar ($l = 2$) component is about 29~\%. The fraction of reconstructed magnetic energy showing up in modes with $m = 0$ is 27~\% of the total energy (Table~\ref{table:ZDIanalysis}). The magnetic ZDI analysis shows that the maximum field strength on the map is 32~G and the average field strength is 7.4~G.

\subsection{Longitudinal magnetic field, line activity indicators and radial velocity}

\begin{figure}[!hc]
      \centering
      \includegraphics[width=10cm]{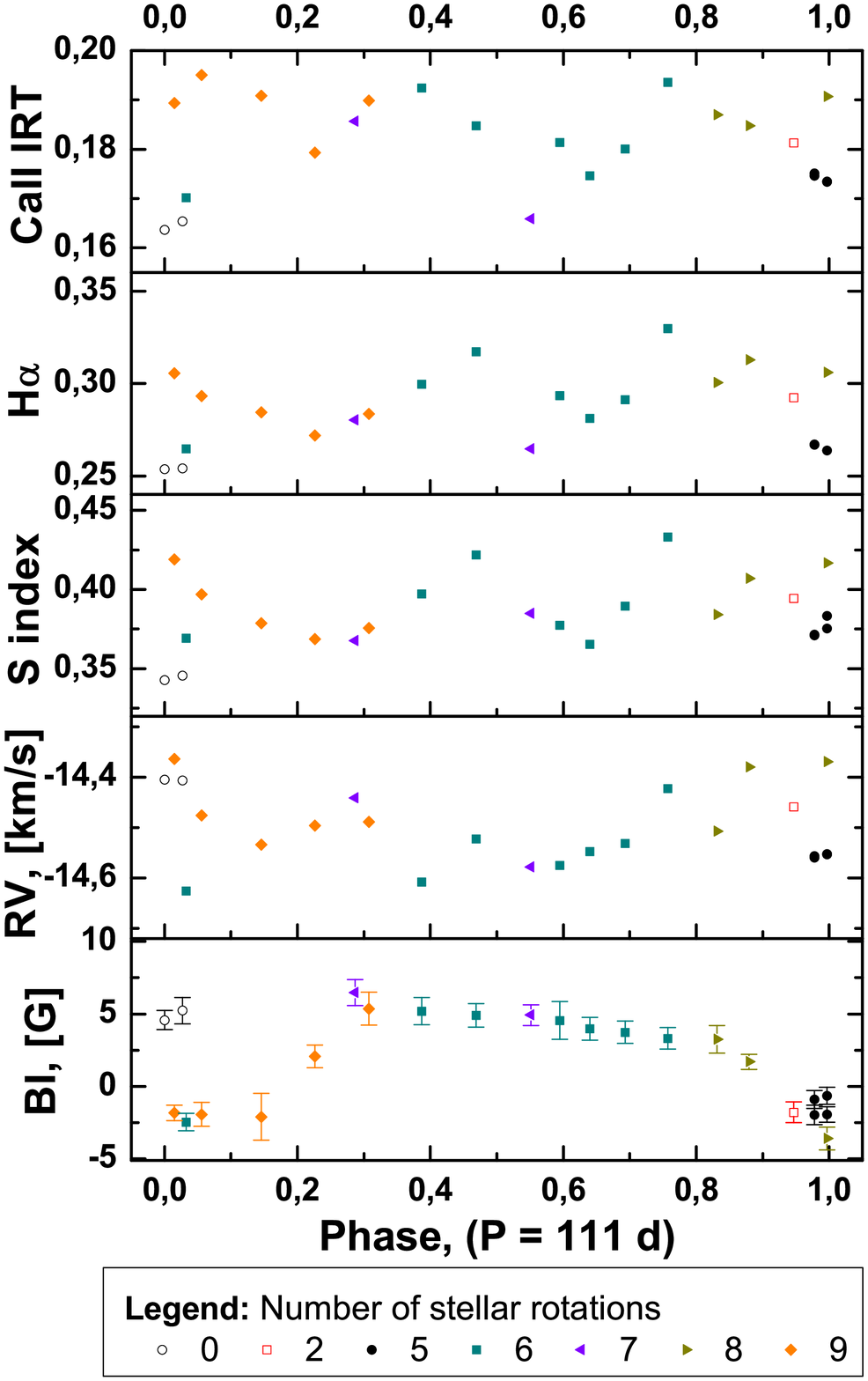}
      \caption{From top to bottom: time variability of Ca\,{\sc ii} IRT, H$\alpha$, S-index, radial velocity, and $B_{l}$ for the period from April 2008 to February 2011 for the giant 37 Com. The legend gives the number of the stellar rotations, with each rotation indicated by a different symbol and color.}
   \label{fig:Bl}
   \end{figure}

The variability of the longitudinal magnetic field $B_{l}$, line activity indicators (Ca\,{\sc ii} IRT, H$\alpha$ and S-index) and the radial velocity (RV) of the giant 37 Com are shown in Fig.~\ref{fig:Bl}. This figure represents all the data presented in this study (from April 2008 to February 2011; Table~\ref{table:1}) as a function of the rotational phases (with a rotation period of 111 days) of the observations, which are listed in column 4 of Table~\ref{table:1}. The open symbols (phase 0 and 2 in the table) denote the data that were discarded for the ZDI map reconstruction -- two black circles from stellar rotation 0 (two spectra from 2008) and one red square from stellar rotation 2 (one spectrum from 2009). The single point from 2009 does not stand out from the general trends of variability for all the line activity indicators and $B_{l}$. On the contrary, the two points from 2008 show positive signs of $B_{l}$, while all of our other measurements are negative, indicating that, during 2008, the negative spot, which is visible near phase 0.0 in the ZDI map, probably did not exist at that location, or that it has moved there since then.

The significant variations of the magnetic field $B_{l}$ are of the order of several gauss. Except for the two points from 2008, the rest of our data show that there are two sign reversals of the $B_{l}$ curve during a given rotational cycle -- the first one being around phase 0.2 changing sign from negative to positive, and the second one around phase 0.9, changing back sign from positive to negative.

Figure~\ref{fig:Bl} shows that there is a good correlation of the time variability among the three line activity indicators Ca\,{\sc ii} IRT, H$\alpha$, and S-index and the RV, but the correlation with $B_{l}$, which shows a smooth, single-peaked curve, is not very strong. Probably, this discrepancy suggests the presence of a small-scale magnetic structure, spatially unresolved in our ZDI model.

Our measurements of the RV are in good agreement with the data given in the catalogs of de Medeiros \& Mayor (1999) and de Bruijne \& Eilers (2012).


\section{37 Com: CNO abundances and $^{12}$C/$^{13}$C ratio}
\label{sec:abundances}

Atmospheric parameters of 37 Com were determined by McWilliam (1990), Drake \& Lambert (1994), and Adamczak \& Lambert (2013). The values derived by these authors are presented in Table~\ref{tab:atmpar}. To check these parameters, we calculated a grid of synthetic spectra in a wide range of effective temperatures and surface gravities ($T_{\rm eff}/\log g$) in the spectral region 6080 -- 6180~\AA\ . This contains lines of neutral elements with different excitation potentials and different equivalent widths and two ionised iron lines, Fe\,{\sc ii} at $\lambda$6084.1~\AA\ and $\lambda$6149.2~\AA\ , which are sensitive to surface gravity.

Synthetic spectra were calculated using the local thermodynamic equilibrium (LTE) model atmospheres of Kurucz (1993) and the current version of the spectral analysis code {\sc moog} (Sneden 1973). The VALD atomic data base (Kupka et al. 1999) was used to create the line lists. Synthetic spectra were calculated with different values of $v\sin i$ with a step of 0.5~km\,s$^{-1}$. A macroturbulence velocity of 3~km\,s$^{-1}$  was adopted as a typical value for the red giant stars (Fekel 1997).

Our tests showed that a synthetic spectrum calculated with the atmospheric model from Drake \& Lambert (1994) ($T_{\rm eff}=4625$~K, $\log g=2.3$, and [Fe/H]\,=\,--0.05) fits the observed one well. However, a low microturbulent velocity ($\xi_{\rm m}=1.0$~km\,s$^{-1}$) does not permit us to fit strong and weak iron lines with the same iron abundance. The best fit was achieved using $\xi_{\rm m}=1.5$~km\,s$^{-1}$, the value typical for field K giant stars. If we adopt the low metallicity ([Fe/H] = --0.53) derived by Adamczak \& Lambert (2013), we are unable to match weak Fe\,{\sc i} lines (which are relatively insensitive to variations of microturbulent velocity). The low metallicity that they derived may be a consequence of their large microturbulent velocity, $\xi_{\rm m}=2.8$~km\,s$^{-1}$, which increases the equivalent widths of stronger lines. Synthetic spectra calculated with the high value of $T_{\rm eff}=4850$~K derived by McWilliam (1990) do not fit the observed spectrum. The adopted atmospheric parameters are shown in Table~\ref{tab:atmpar}.

\begin{table}[!hc]
\centering
\caption{37~Com: Atmospheric parameters and metallicity.}
\label{tab:atmpar}
\begin{tabular}{cccccc}\hline
$T_{\rm eff}$ & $\log g$ & $\xi_{\rm m}$ & [Fe/H] & Ref  \\
       K      &          & km\,s$^{-1}$  &        &      \\
\hline
4\,625 & 2.3 & 1.0 & $-$0.05 & (1)       \\
4\,850 & 2.8 & 2.9 &   +0.14 & (2)       \\
4\,610 & 2.5 & 2.8 & $-$0.53 & (3)       \\
4\,625 & 2.3 & 1.5 & $-$0.05 & This work \\
\hline
\end{tabular}
{\par (1) Drake \& Lambert (1994),
\par (2) McWilliam (1990),
\par (3) Adamczak \& Lambert (2013)}
\end{table}

\begin{figure}[!hc]
   \centering
   \includegraphics[width=9.5cm]{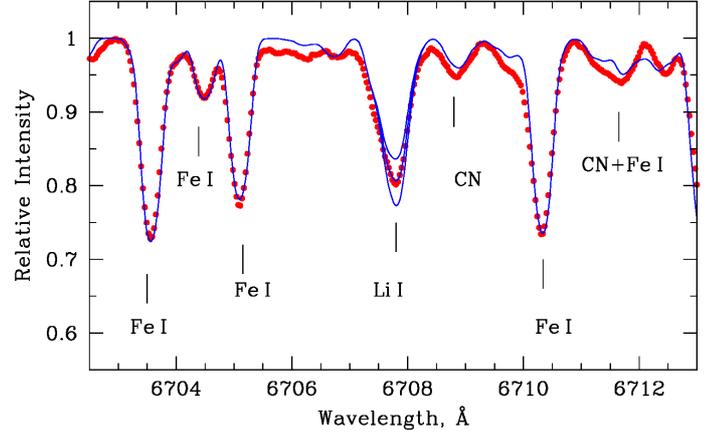}
   \caption{Observed (dotted red line) and synthetic (solid lines) spectra in the region of the Li\,{\sc i} resonance doublet. Synthetic spectra were calculated for lithium abundances $\log\varepsilon{\rm (Li)}=1.13$, 1.23, and 1.33.}
\label{fig:Li}
\end{figure}

It is well known that many weak G-band giant stars have high Li abundance (new determinations and a summary of other works are presented in Palacios et al. 2012). We derived the Li abundance of 37 Com by spectrum synthesis of the Li\,{\sc i} resonance doublet at $\lambda$6707.8~\AA. The wavelengths and oscillator strengths for individual hyperfine and isotopic components of the lithium lines were taken from Smith et al. (1998) and Hobbs et al. (1999). A solar $^6$Li/$^7$Li isotopic ratio ($^6$Li/$^7$Li\,=\,0.081; Anders \& Grevesse 1989) was adopted in synthetic spectrum calculations. We derived a lithium abundance $\log\varepsilon{\rm (Li)} =1.23 \pm 0.15$ (where $\log\varepsilon{\rm (Li)}= \log (N_{\rm Li}/N_{\rm H})+12$), a value that is significantly higher than the mean Li abundance in the atmospheres of G and K giants ($\langle\log\varepsilon{\rm (Li)}\rangle\sim +0.1$), obtained by Brown et al. (1989). Assuming typical errors in the effective temperature, gravity and microturbulent velocity of $\sigma(T_{\rm eff})=\pm 100$~K, $\sigma(\log g)= \pm 0.2$ and $\sigma(\xi_{\rm m})=\pm 0.3$~km\,s$^{-1}$, we estimated the uncertainty on the Li abundance. Our calculations showed that the uncertainty in the effective temperature $\Delta T_{\rm eff} =+100$~K results in an increase of the Li abundance of $\Delta\log\varepsilon({\rm Li})=+0.15$~dex, whereas the estimated uncertainties in $\log g$ and $\xi_{\rm m}$ (0.2 and 0.3~km\,s$^{-1}$, respectively) do not affect the Li abundance in 37 Com. A comparison of the observed and synthetic spectra is shown in Fig.~\ref{fig:Li}.

\begin{table*}[!htc]
\centering
\begin{center}
\caption{37~Com: Light elements and Li abundances. The last column provides the $^{12}$C/$^{13}$C isotopic ratios.}
\begin{tabular}{cccccc}
\hline
$\log\varepsilon$(Li) & $\log\varepsilon$(C) & $\log\varepsilon$(N) & $\log\varepsilon$(O) & $^{12}$C/$^{13}$C & Ref \\
\hline
+1.23$\pm$0.15 & $\le$\,7.02          & 9.57$\pm$0.14 & 8.83$\pm$0.11 & 4.2$\pm$0.2 & (1) \\
+1.23$\pm$0.15 & 6.82$\pm$0.25$^\ast$ & 9.90$\pm$0.14 & 8.83$\pm$0.11 & 4.2$\pm$0.2 & (1) \\
+1.50$\pm$0.25 & 6.82$\pm$0.25        & 9.77$\pm$0.3  & 8.72$\pm$0.11 & 3.5$\pm$0.5 & (2) \\
\hline
\end{tabular}
{\par $^\ast$ Value from Adamczak \& Lambert (2013),
\par (1) This work,
\par (2) Adamczak \& Lambert (2013)}
\label{tab:LiC1213}
\end{center}
\end{table*}

Determination of CNO abundances and the $^{12}$C/$^{13}$C ratio is important to understand the evolutionary status of 37 Com. The oxygen abundance was determined using the forbidden oxygen line at $\lambda$6300.304~\AA. To determine the carbon abundance, we analysed three spectral regions containing the molecules C$_2$ and CH: $\sim\lambda$4360 (CH), $\sim\lambda$5086 (C$_2$), and $\sim\lambda$5635 (C$_2$). The line lists used in this study are the same as those described by Drake \& Pereira (2008), with the exception of the dissociation energy of the CN molecule which, in this paper, was taken to be equal to 7.75~eV, based on experimental determinations and theoretical calculations (Costes et al. 1990, Huang et al. 1992, Pradhan et al. 1994). We determined only the upper limit of the carbon abundance, $\log\varepsilon{\rm (C)}\le 7.02$, for this weak G-band star. Adamczak \& Lambert (2013), using similar atmosphere parameters, derived a carbon abundance $\log\varepsilon{\rm (C)} =6.82$.

The eventuality of contamination of the [O\,{\sc i}] $\lambda$6300.304~\AA\ line by telluric O$_2$ lines and $^{12}$CN and $^{13}$CN lines at  $\sim\!\lambda$8004~\AA\ by telluric H$_2$O lines was checked using a hot star spectrum. We verified that neither the [O\,{\sc i}] line at $\lambda$6300.304~\AA\ used for the oxygen abundance determination, nor the $^{13}$CN feature used for carbon isotope ratio determination are blended with telluric lines in our spectrum. Derived light element abundances, as well as $^{12}$C/$^{13}$C isotopic ratios, are shown in Table~\ref{tab:LiC1213}.

\begin{figure}[!hc]
   \centering
   \includegraphics[width=9.5cm]{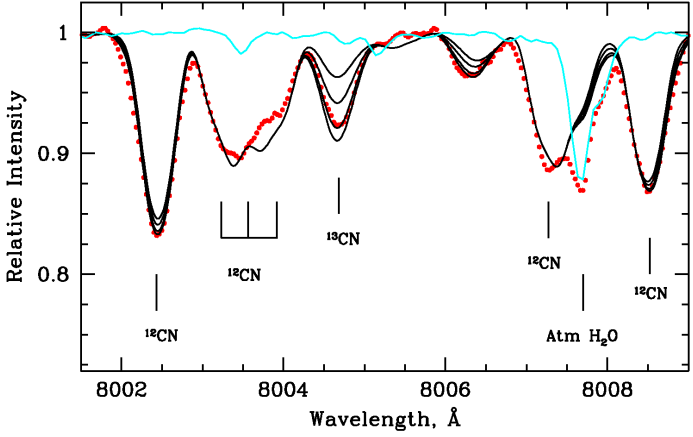}
   \caption{Observed (dotted red line) and synthetic (solid lines) spectra in the region around the $^{12}$CN and $^{13}$CN lines at $\sim\lambda$8004~\AA. Synthetic spectra were calculated for four $^{12}$C/$^{13}$C ratios, 3.6, 4.2, 6.0, and 10.0. The light blue solid line shows the spectrum of a hot star used to map the telluric H$_2$O lines.}
\label{fig:C12C13}
\end{figure}

The error in the derived $^{12}$C/$^{13}$C ratio is mainly due to uncertainties in the observed spectrum, such as possible contamination by unidentified atomic or molecular lines or uncertainties in the continuum placement. Comparison of the observed and synthetic spectra is shown in Fig.~\ref{fig:C12C13}.

The nitrogen abundance was determined using the lines of the CN molecule in the $\sim$8000~\AA\ spectral region. Intensities of the CN lines depend on abundances of carbon, nitrogen, and oxygen (since abundances of the CNO elements are interdependent, because of the association of carbon and oxygen in CO molecules in the atmospheres of cool giants). In the nitrogen abundance determination, we used the oxygen abundance value in this paper and carbon abundances presented by Adamczak \& Lambert (2013) and the upper limit derived in this paper.

Assuming the same typical values of uncertainties in the effective temperature and surface gravity ($\sigma(T_{\rm eff})=\pm$100~K and $\sigma(\log g) = \pm 0.2$), we estimated the errors in the CNO abundances. Derived CNO abundances are weakly sensitive to the variations of the microturbulent velocity since weak lines were used to determine them. Derived light element abundances as well as $^{12}$C/$^{13}$C isotopic ratios are shown in Table~\ref{tab:LiC1213}.

\section{Mass and evolutionary status of the giant 37 Com}
\label{sec:evolution}

\subsection{The peculiar abundances of 37 Com and the need for dedicated models}

The giant 37 Com is identified among the rare class of weak G-band stars (hereafter wGb). wGb stars are G and K giants whose spectra show very weak or no absorption in the G-band of the molecule CH at 4300~\AA. Already proposed as this type of star by Lambert \& Ries (1981), a new analysis of the fundamental parameters and surface abundances of 37 Com was performed by Adamczak \& Lambert (2013). Their analysis revealed a strong deficiency in carbon together with high nitrogen and lithium abundances. This is consistent  with the very low value that we derived for the carbon isotopic ratio and the upper limit found for the carbon abundance, as indicated in the previous section. Such an abundance pattern is very difficult to reconcile with a self-pollution scenario associated with non-standard mixing processes within the star itself. Indeed, the CNO abundance pattern of wGb stars is characteristic of CNO cycle nucleosynthesis that actually occurs in the hydrogen burning shell of intermediate-mass red giant stars. However there are currently two major arguments against a self-pollution scenario in this type of star. First, to explain the surface CNO abundance pattern would require an extremely efficient transport process to connect the convective envelope to the bulk of the energy generation shell where the hot CNO cycle occurs. This would strongly alter the evolution of the star in a manner that is not evidenced from observations. Secondly, the region where the CNO cycle occurs is too hot for lithium to survive proton capture, therefore any mixing that would explain the surface CNO abundance pattern would be incompatible with A(Li) = 1.23 dex. Another possibility is that wGb stars, including 37 Com, formed out of CNO processed material or were polluted by such material in their infancy, as suggested by Palacios et al. (2016). In that case, the peculiar initial chemical composition of the plasma should affect the evolutionary path of the star, and it is thus necessary to use dedicated modeling to evaluate its mass and evolutionary status, assuming that the star was initially C-poor and N-rich.

We computed tailor-made rotating models of different masses, initial chemical composition, and rotation rates using the code STAREVOL (see Siess 2006, Decressin et al. 2009). The assumptions of the models are described in Table~\ref{tab:models}. We assumed a reference solar chemical mixture from Asplund et al. (2005). Rotation-induced mixing is as described by Decressin et al. (2009), core overshooting is computed as described by Charbonnel \& Lagarde (2010), and mass loss is modeled according to Reimers' (1975) formula. The initial rotation velocities adopted here are within the distribution of expected velocities for late B-type stars and lead to $P_{\rm rot}$ between 110 and 180 days in the temperature range 4500-4700~K, which corresponds to the temperature of 37 Com.

\begin{table*}
\caption{Dedicated stellar evolution models computed from the PMS to the early-AGB stage. All models are computed using Asplund et al. (2005) as a reference for the solar abundance mixture, Z$_\odot$ = 0.012294~dex, A(C)$_\odot$ = 8.43, A(N)$_\odot$ = 7.83, A(O)$_\odot$ = 8.69.}
\centering
\begin{tabular}{c c c c c c c}
\hline
   M        & $< \upsilon_{ZAMS} > $ & Z & [Fe/H] & [C/Fe]$_{ini}$ & [N/Fe]$_{ini}$  & [O/Fe]$_{ini}$ \\
($M_\odot$) &    km\,s$^{-1}$        &   & (dex)  &  (dex)         & (dex)           &  (dex)         \\
\hline
\hline
5.0 & 60            & 0.011  & -0.05 & 0     & 0    & 0    \\
5.8 & 60            & 0.0637 & -0.05 & -1.56 & 1.99 & 0.08 \\
6.2 & 60            & 0.0637 & -0.05 & -1.56 & 1.99 & 0.08 \\
6.5 & 60,80,110,120 & 0.0637 & -0.05 & -1.56 & 1.99 & 0.08 \\
6.6 & 110           & 0.0637 & -0.05 & -1.56 & 1.99 & 0.08 \\
7.0 & 110           & 0.0637 & -0.05 & -1.56 & 1.99 & 0.08 \\
\hline
\end{tabular}
\label{tab:models}
\end{table*}

\subsection{Model predictions : Mass, evolution status, rotation, and Rossby number}

\begin{figure}
   \includegraphics[width=0.45\textwidth]{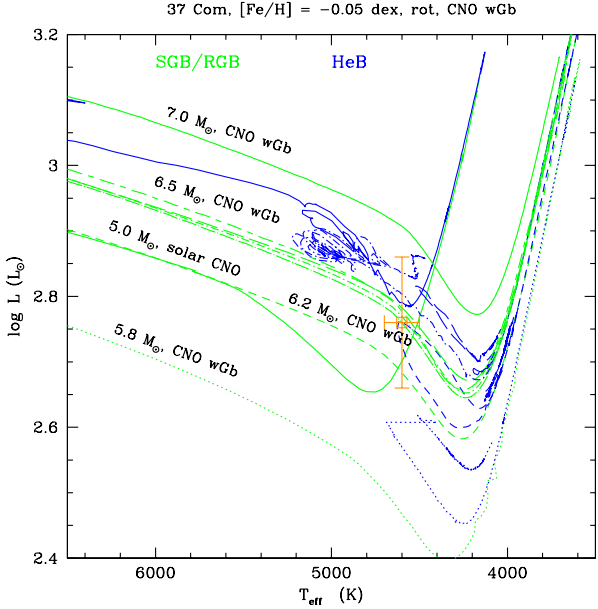}
   \caption{Hertzsprung-Russell diagram including the position of 37 Com (using the latest version of Hipparcos parallaxes) and the rotating stellar evolution models described in Table~\ref{tab:models} and labeled on the figure. The label 6.5~$M_\odot$, CNO wGb refers to the set of 4 tracks shown in broken lines right below the label, each of them computed with a different rotational velocity on the ZAMS: 60 km\,s$^{-1}$, 80 km\,s$^{-1}$, 110 km\,s$^{-1}$ and 120 km\,s$^{-1}$. The 5~$M_\odot$ model with solar CNO abundances is the solid line in the middle of the plot. In the colour version of this figure, the green parts of the tracks correspond to the subgiant/red giant branch phases, and the blue parts are associated with the core He-burning phase.}
\label{fig:hrd}
\end{figure}

Figure~\ref{fig:hrd} shows the position of 37 Com in the Hertzsprung-Russell diagram together with the models described in Table~\ref{tab:models}. The luminosity is computed using the classical formula $log_{10}(L/L_\odot) = 0.4*\left( 5 log_{10}(d_{pc}) - 5 + 4.74 - m_V - BC_V + A_V \right)$, where we adopted $BC_V = -0.528$ from Flower (1996) , $m_V = 4.88$ from the Hipparcos catalog (Perryman et al. 1997), $d_{pc} = 202.43 ^{+25.36}_{-20.28}$~pc after the Hipparcos parallax (Van Leeuwen et al. 2007), and $A_V$ = 0.0307 from the NASA/IPAC IRSA DUST application \footnote{http:\/\/irsa.ipac.caltech.edu/applications\/DUST\/}. We adopt ($T_{\rm eff}$, log($L/L_\odot$) $\equiv$ (4600 $\pm$ 100~K, 2.78$^{+0.1}_{-0.09}$)). The models computed with modified CNO abundances appear redder and less luminous that the model computed with a solar mixture of these elements. This is due to the fact that for the same iron content [Fe/H], modifying the C and N abundances leads to an overall metallicity $Z$ larger by a factor of 6, thus changing the opacity and shifting the evolutionary tracks down in luminosity and temperature. The position of 37 Com in the HR diagram is compatible with a 5.0~$M_\odot$ progenitor at [Fe/H] = -0.05~dex with a standard solar mixture of CNO elements in the Hertzsprung gap or at the red clump (core He-burning phase), and with a 6.5~$M_\odot$ progenitor at [Fe/H] = -0.05~dex with a modified C,N,O mixture in accordance with the values estimated by Adamczak \& Lambert (2013) (e.g. A(C)$_\odot$ = 6.82~dex, A(N)$_\odot$ = 9.77~dex, A(O)$_\odot$ = 8.72~dex) in the Hertzsprung gap close to the base of the red giant branch (RGB). This diagram reveals the importance of taking into account the peculiar abundances of 37 Com to determine its mass and evolutionary status, but is not conclusive in itself.

\begin{figure}
   \includegraphics[width=0.40\textwidth]{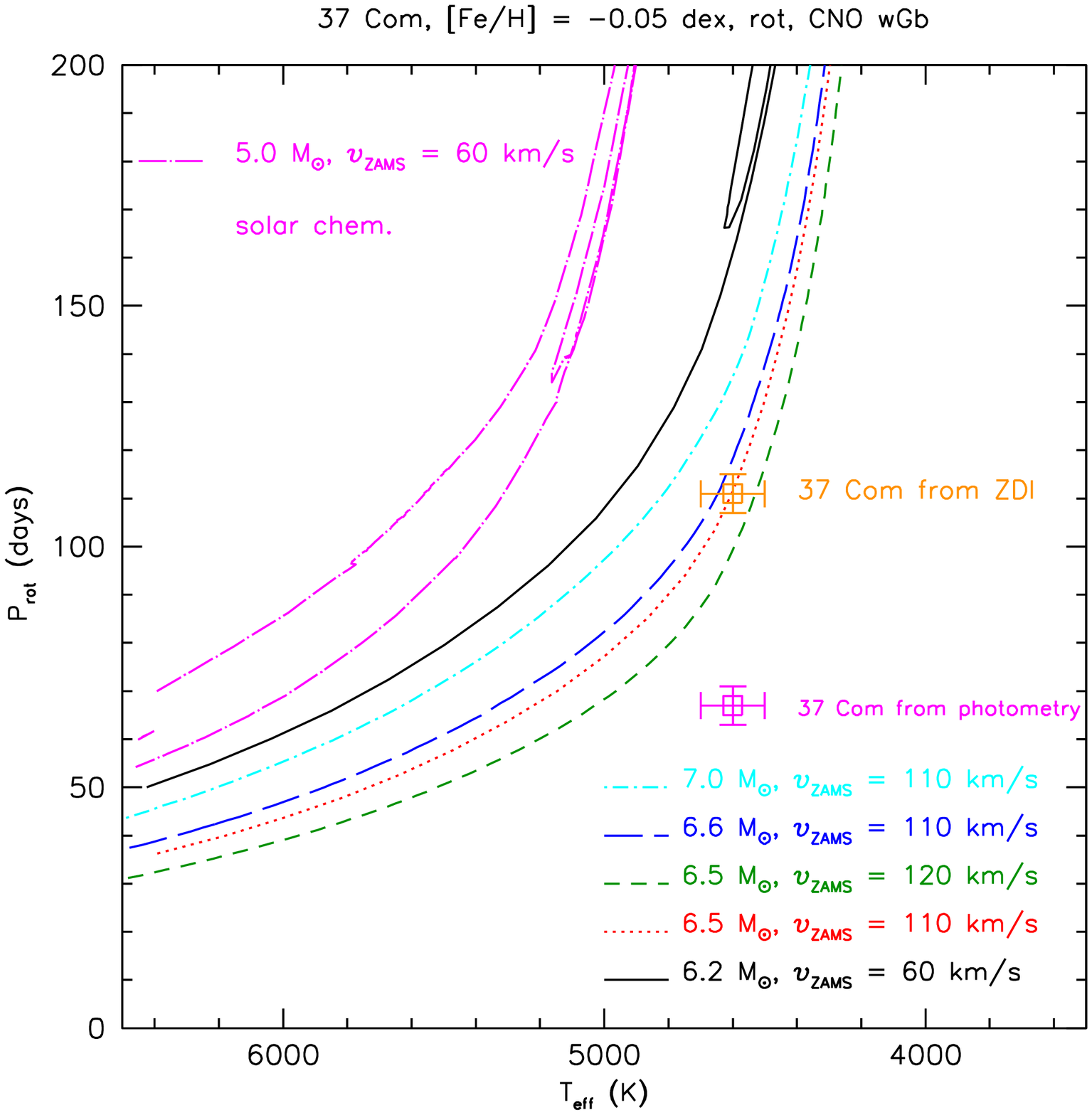}\\
   \includegraphics[width=0.40\textwidth]{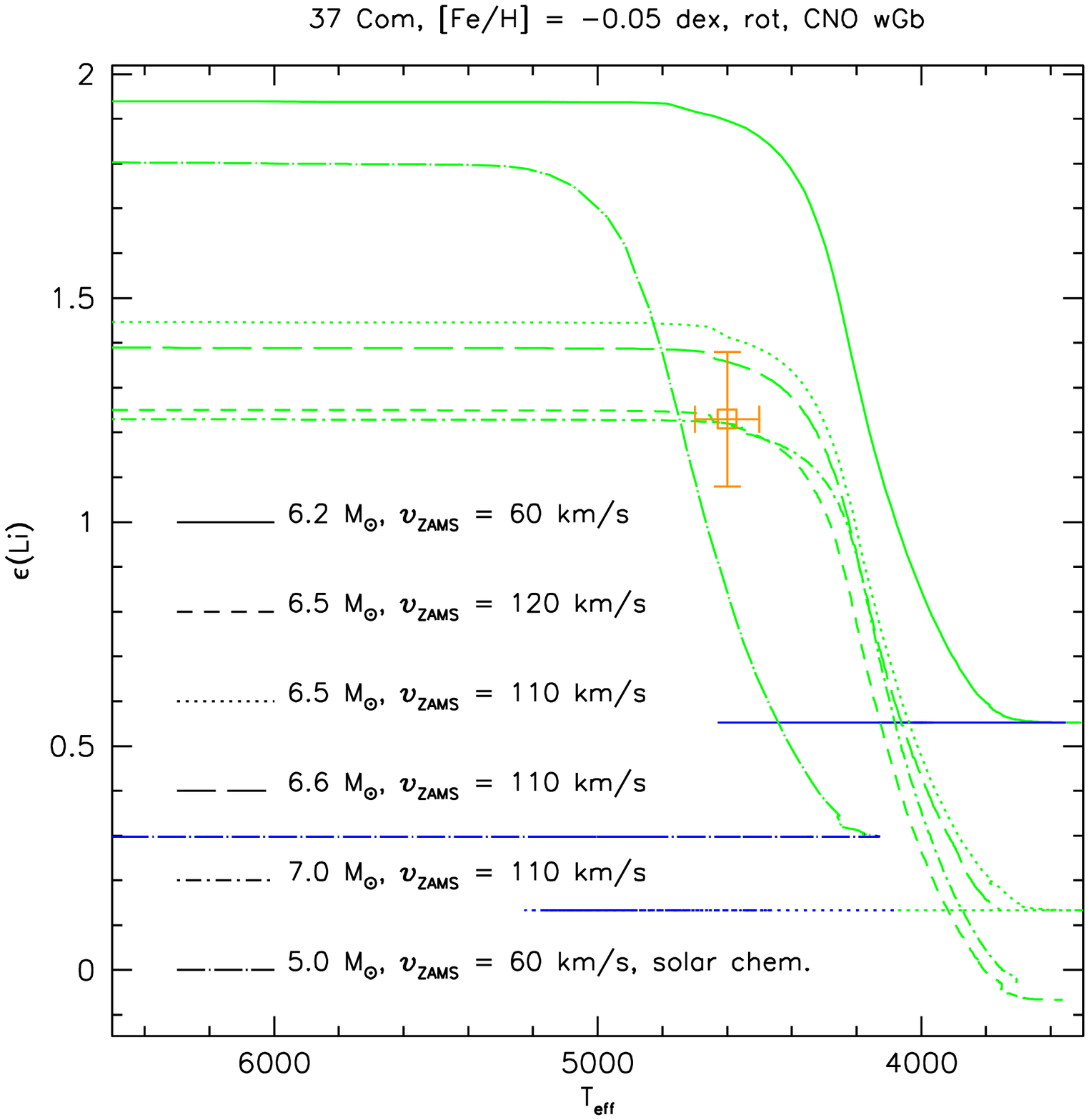}\\
   \includegraphics[width=0.40\textwidth]{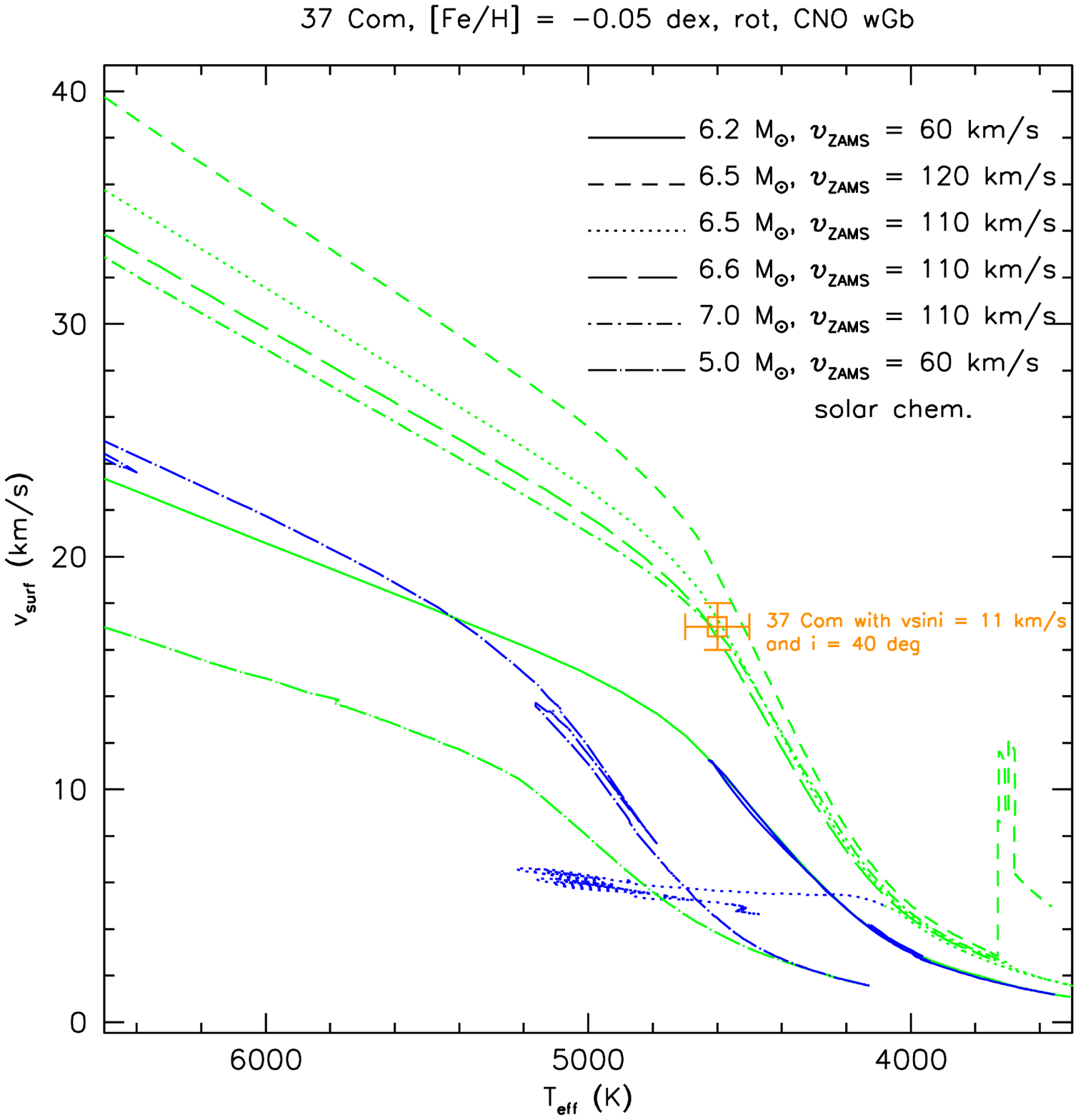}
   \caption{Evolution as a function of $T_{\rm eff}$ for $P_{\rm rot}$, $\epsilon (Li)$ and $v_{\rm surf}$ for some models described in Table~\ref{tab:models}, as labeled on the plots. The position of 37 Com is reported on each panel. Its lithium abundance is that determined in the present study : A(Li) = 1.23 $\pm$ 0.15. The error bar on the effective temperature is about 100~K. In the upper panel, we also show the photometric period, which is not compatible with the other indicators, contrary to the one derived from ZDI and adopted in the present study. In the middle and lower panels, the colours are the same as in Fig.~\ref{fig:hrd}.}
\label{fig:modcons}
\end{figure}

However, when we combine the observational constraints available in terms of $P_{\rm rot}$, Li surface abundance, $^{12}$C/$^{13}$C abundance ratio, $T_{\rm eff}$, and luminosity, the degeneracy can be lifted as demonstrated in Fig.~\ref{fig:modcons}. First, the 5~$M_\odot$ model with solar scaled initial abundances is clearly ruled out because the data point does not intersect the associated track in any of the planes represented. Second, the tracks associated with the 6.2~$M_\odot$ model at moderate initial velocity are also incompatible with observations (too much lithium, and too slow rotation at the $T_{\rm eff}$ of 37 Com). The best fit for the lithium is given by the 6.5~$M_\odot$ and 7~$M_\odot$ models with $v_{\rm ZAMS} = 120$ and $110$ km\,s$^{-1}$, respectively, while the rotation (both period and $v_{\rm surf}$) is best fit by the 6.5~$M_\odot$ with $v_{\rm ZAMS} = 110$ km\,s$^{-1}$. Hence, from all these indicators, 37 Com appears to be a star that is crossing the Hertzsprung gap (i.e. a young red giant), of about 6.5~$M_\odot$ that had an initial rotation rate of about 120~km\,s$^{-1}$ on the zero age main sequence, and initially low carbon and high nitrogen content, consistent with other wGb stars. Accounting for all the available observational constraints, 37 Com is compatible with a very narrow mass range of 6.5 $\pm 0.1$~$M_\odot$, and $v_{\rm ZAMS}$ in the range 110 to 120 km\,s$^{-1}$.

\begin{figure}
   \centering
   \includegraphics[width=0.45\textwidth]{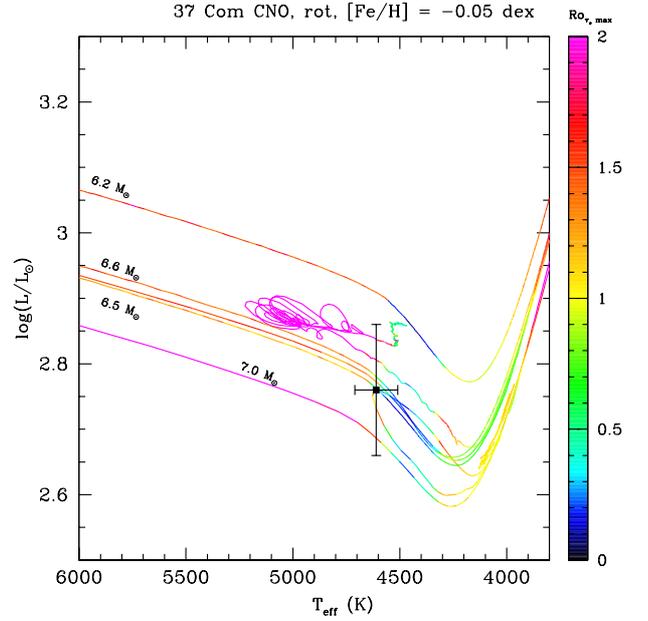}
   \caption{Evolution of the Rossby number (colour coded) along the tracks in the HRD, and position of 37 Com. The error bar on the effective temperature is about 100~K.}
\label{fig:Rossby}
\end{figure}

The evolution of Rossby number (defined as $Ro = P_{\rm rot}/ \tau _{c}$, with $\tau_c$ the maximum convective turnover timescale in the envelope) of 37 Com as a function of evolution is shown in Fig~\ref{fig:Rossby}, as a third dimension in the form of a colour coding of the tracks. The figure presents a zoom around the region of the RGB and the clump for tracks of rotating stars with initial carbon depletion and nitrogen enhancement as expected for the wGb stars. This figure emphasizes how fast the Rossby number changes, when approaching the base of the RGB, due to the growth of the convective envelope, providing more favorable conditions for dynamo action. As can be seen from the colour coding, the modeled Rossby number of 37 Com is of about $0.7^{+1.8}_{-0.4}$ with a large uncertainty owing to the error bars on the temperature and luminosity. These uncertainties simply represent the maximum and minimum values expected within the uncertainty box in the HR diagram.

\section{Discussion}

Two main hypotheses have been proposed to explain the presence of magnetic fields in single late-type giants. The first one is the existence of a stellar dynamo, which continuously generates the magnetic fields, because these stars have extended convective envelopes. The second hypothesis is that a fraction of cool giants could be descendants of Ap stars. It is believed that the magnetism in Ap stars is due to fossil fields (Braithwaite \& Spruit 2015). On the main sequence, these fields are stable with time and show a roughly dipolar structure of their large-scale surface magnetic fields (Wade et al. 2000 a, b, Auri\`ere et al. 2007, Kochukhov \& Wade 2010, Silvester et al. 2012). After the main sequence, the star enters the Hertzsprung gap, and the fossil field may still survive in the stellar interior and interact with the convective upper layers (Strugarek et al. 2011).

Being a member of the sample of Auri\`ere et al. (2015), it has already been proposed that the giant 37 Com owes its magnetic characteristics to the activity of a dynamo. These authors report several relations between the rotational periods $P_{\rm rot}$, magnetic field $|B_{l}|_{max}$, $L_{x}$, Rossby number and S-index for giants, which have dynamo-generated magnetic fields. The activity proxies and magnetic quantities estimated for 37 Com follow the trends reported by Auri\`ere et al. (2015). There are several giants, which do not fit these trends, for example EK Eri and $\beta$ Ceti. The detailed studies of the magnetic field structure and origin of these two giants, EK Eri (Auri\`ere et al. 2011) and $\beta$ Ceti (Tsvetkova et al. 2013), show that they could be descendants of magnetic Ap stars and host fossil fields.

Some of the properties of the giants 37 Com, EK Eri, $\beta$ Ceti and, in addition, V390 Aur (another giant with dynamo action; Konstantinova-Antova et al. 2012), are summarized in Table~\ref{table:fossildynamo}. It should be noted that there are some differences between the two groups of stars -- 37 Com and V390 Aur belong to the first group; they rotate faster and are more active than the second group, in the sense of a higher X-ray luminosity (without a clearly distinctive S-index).

\begin{table*}
\centering
\begin{center}
\caption{Properties of the giants with dynamo activity (37 Com, V390 Aur) and for giants with probable fossil fields ($\beta$ Ceti, EK Eri) (from Auri\`ere et al. 2015 and references therein).}
\label{table:fossildynamo}      
\centering                          
\begin{tabular}{c c c c c c c c c}        
\hline\hline               
Name & Sp. type & Evolution & $v \sin i$     & $P_{\rm rot}$ & $\mid B_{l} \mid_{max}$ & S-index & Ro &    $L_{x}$             \\
     &          & phase     & [km\,s$^{-1}$] &  [days]       &          [G]            &         &    & $10^{27}$ erg\,s$^{-1}$\\
\hline\hline
\multicolumn{9}{c}{Dynamo action} \\ \hline
37 Com   & G9~III & H-gap    & 11 & 111 & 6.5 & 0.368 & 0.70 & 5200 \\
V390 Aur & G8~III & Base RGB & 29 & 9.8 & 13  & 0.681 & 0.04 & 5040 \\ \hline
\multicolumn{9}{c}{Fossil field origin} \\ \hline
$\beta$ Ceti & K0~III    & He-burning & 3.5  & 215   & 10 & 0.236 & 0.93 & 1585 \\
EK Eri       & G8~III-IV & Base RGB   & 1.0  & 308.8 & 99 & 0.501 & 1.30 & 1000 \\
\hline\hline
\end{tabular}
\\
\end{center}
\end{table*}

\begin{table*}
\centering
\begin{center}
\caption{Magnetic characteristics of 37 Com (present study), V390 Aur, $\beta$ Ceti, and EK Eri.}
\label{table:ZDIanalysis}      
\centering                          
\begin{tabular}{c c c c c c}        
\hline\hline               
     & pol. comp.    & dipole comp.  & quad. comp.   & oct. comp.    & axi. comp.    \\
Name & (\% tot)      & (\% pol)      & (\% pol)      & (\% pol)      & (\% tot)      \\
\hline\hline
\multicolumn{6}{c}{Dynamo action} \\ \hline
37 Com   & 88.0 & 22.2 & 29.2 & 26.7 & 27.3 \\
V390 Aur & 87.0 & 25.5 & 15.7 & 12.5 & 46.7 \\ \hline
\multicolumn{6}{c}{Fossil field origin} \\ \hline
$\beta$ Ceti 2010      & 96.7   & 83.2 & 20.8 & 6.2 & 77.1 \\
$\beta$ Ceti 2011/2012 & 96.5   & 85.4 & 11.3 & 4.0 & 74.4 \\
EK Eri                 & $>$ 99 & 91.6 & 5.6  & 2.8 & 80.6 \\
\hline\hline
\end{tabular}

Note: 2nd to 5th columns list the fraction of the large-scale magnetic energy reconstructed in the poloidal field component, the fraction of the poloidal magnetic energy stored in the dipolar ($l=1$), quadrupolar ($l=2$) and octopolar ($l=3$) components, and the fraction of the energy stored in the axisymmetric component ($m=0$).
\\
\end{center}
\end{table*}

The level of complexity of the surface magnetic structures for 37 Com, V390 Aur, EK Eri, and $\beta$ Ceti, as revealed by their ZDI maps, is different. The structures of EK Eri and $\beta$ Ceti are quite simple (as expected for stars with fossil fields). On the other hand, the giant V390 Aur shows very complex structures. The ZDI map of 37 Com, which we present in this study (Fig.~\ref{fig:37ComZDImap}), is not as simple as for EK Eri and $\beta$ Ceti, nor is it as complex as for the giant V390 Aur. The different $v \sin i$ values, leading to a different spatial resolution of the ZDI method, could be partially responsible for these differences. We note however that even the quadrupoles and octupoles of EK Eri and $\beta$ Ceti store a small fraction of their magnetic energy, while such low-order magnetic components should be resolved at their low projected rotational velocity (Morin et al. 2010).

Following the analysis of the field topology presented in Table~\ref{table:ZDIanalysis}, the poloidal component of the magnetic field is dominant for all the giants. The difference here is that for EK Eri and $\beta$ Ceti the percentage of the toroidal component of the magnetic field is close to zero while, for the giants 37 Com and V390 Aur, the toroidal component represents a significant percentage of the magnetic energy, as previously noticed for a number of stars with a dynamo field (See et al. 2015). In this case again, possible technical limits of ZDI at low $v \sin i$ are unlikely to be the reason for the non-detection of toroidal magnetic components for EK Eri and $\beta$ Ceti. Here, a comparison with main-sequence stars is helpful, as it shows that strong poloidal fields are routinely reconstructed for cool, low $v \sin i$ dwarfs (Morgenthaler et al. 2012, Folsom et al. 2016). It is therefore most likely that the absence of significant toroidal components for these slowly rotating and strongly magnetic cool giants is genuine, and can be interpreted as another hint in favour of a fossil origin of their surface fields, as discussed by Auri\`ere et al. (2011), Tsvetkova et al. (2013), and Borisova et al. (2016).

One additional constraint is the measurement of the differential rotation parameters. As given in Sect.~\ref{sec:diffrotparameters} for 37 Com, we derive $\Omega_{\rm eq} = 0.061 \pm 0.001$~rad/d and $d \Omega = 0.009 \pm 0.001$~rad/d. These parameters are also available for V390 Aur (Konstantinova-Antova et al. 2012): $\Omega_{\rm eq} = 0.652 \pm 0.002$~rad/d and $d \Omega = 0.048 \pm 0.007$~rad/d. Whereas, the measurements for the giants EK Eri and $\beta$ Ceti were inconclusive (but presumably very weak in connection with their long-lived field geometries that would have been progressively modified otherwise). The weak surface shear measured for 37 Com ($d \Omega$ about 1/5 of solar surface differential rotation) seems to be consistent with a large-scale field evolving over several years. This needs to be confirmed by more observations.

On the other hand, another result of our study is that the surface magnetic field of 37 Com shows some variability with time. This is illustrated by the two positive $B_{l}$ measurement at phase $\phi=0$ in 2008, not consistent with more recent measurements. This is different from the stability observed in $\beta$ Ceti, confirmed on the longer term by the published data for the period 2010-2012 (Tsvetkova et al. 2013) and by new data from 2013 (Tsvetkova et al. 2017, in preparation), and in EK Eri.

The correlation between line activity indicators (H$\alpha$, Ca\,{\sc ii} IRT, S-index), $B_{l}$, and radial velocity measurements could provide one more clue about the surface magnetic structure of a giant. For 37 Com, that correlation is not as strong as the ones for EK Eri and $\beta$ Ceti. This difference may indicate that significant magnetic flux remains hidden in spatially unresolved elements that contribute to generating the chromospheric activity of 37 Com (similarly to Sun-like stars, e.g. Morgenthaler et al. 2012), while such small-scale magnetism is much less pronounced in $\beta$ Ceti and EK Eri.

All the results from this study, and the comparison between 37 Com and the other three well-studied giants, provide evidence that 37 Com is a giant with moderate activity that is generated by dynamo-driven magnetic fields. Our study goes further, trying to understand the nature of that giant. Weak G-band stars are rare objects -- less than 30 objects are known in the Galaxy. This is why, in Sect.~\ref{sec:abundances}, we calculated synthetic spectra to define the atmospheric parameters, metallicity, the abundances of C, N, O, and Li elements. We also determine the carbon isotopic ratio $^{12}$C/$^{13}$C$=4.2 \pm 0.2$ to confirm the peculiar abundances of the giant and its status as a weak G-band star. This value corresponds well to the values of other wGb stars from Sneden et al. (1978). We derived the lithium abundance of 37 Com to be $\log\varepsilon{\rm (Li)} =1.23 \pm 0.15$ and this value is higher than that for normal G and K giants, but in the standard interval of values for Li-rich wGb stars -- more than 50~\% of all wGb stars (Palacios et al. 2012).

In Sect.~\ref{sec:evolution}, we show that 37 Com is a 6.5~$M_{\odot}$ star that is crossing the Hertzsprung gap and close to the base of the RGB, using rotating stellar evolutionary models that were computed with  an initial composition suited to a wGb star. According to the model, the progenitor of 37 Com on the main sequence was a late B-type star and its present period of 111 days (reported in this study) and fast projected rotational velocity ($v \sin i = 11 \pm 1$~km\,s$^{-1}$) are consistent with the distribution of expected periods and velocities.

Since wGb stars are rare objects, they have not been studied well until now. The reasons for their carbon underabundance and lithium enrichment (the latter characteristic is valid for more than 50~\% of the objects) are still not clear. The link (if any) between the peculiar abundances of these stars and the magnetic fields has also never been studied. We plan to collect more observational data to study the magnetic fields (if any) in details for several other wGb stars.

\section{Conclusions}

Spectropolarimetric data in the period April 2008-February 2011 were collected for the giant 37 Com. Indirect proxies of magnetism such as Ca\,{\sc ii} IRT, H$\alpha$ and S-index, and the radial velocity have been measured, showing a good correlation among their variability. The longitudinal magnetic field $B_{l}$ has also been measured (of the order of several gauss) and it yields a poor correlation with the line activity indicators, possibly indicating that there are small-scale unresolved magnetic structures. One ZDI map is reconstructed, displaying complex surface magnetic structure. The ZDI analysis reveals that the poloidal component dominates the magnetic topology, but is divided into several components. In addition, the toroidal component is significant and this rather complex topology is consistent with dynamo operation. The two measurements from 2008, which do not fit the general trend of the variability of $B_{l}$, together with the poor fit between observed and modeled Stokes $V$ profiles for observations from 2008 and 2009, give signs of evolution of the surface magnetic structures. Surface differential rotation parameters are also measured, $\Omega_{\rm eq} = 0.061 \pm 0.001$~rad/d and $d \Omega = 0.009 \pm 0.001$~rad/d, showing that $d \Omega$ is about 1/5 of solar value. The modeled Rossby number is about 0.7. All these results show that the magnetic field of the giant 37 Com is a dynamo-generated one.

In addition, synthetic spectra were computed to define the atmospheric parameters, metallicity, the abundances of C, N, O, and Li elements and the carbon isotopic ratio $^{12}$C/$^{13}$C$=4.2 \pm 0.2$. The results confirm the peculiar abundances of 37 Com and suggest that it can be classified as a weak G-band giant star.

On the basis of state-of-the-art stellar evolutionary models, we have shown that 37 Com is a mildly evolved intermediate-mass star in the Hertzsprung gap, in agreement with the evolutionary status found for most weak G-band stars. According to the model, its mass is 6.5~$M_{\odot}$. It is the first wGb giant with a detected magnetic field, which adds yet another piece to the puzzle of their anomalous surface chemical composition.

We also compared the results of 37 Com with the results from three detailed studied giants. For V390 Aur, the magnetic field is dynamo-generated. For EK Eri and $\beta$ Ceti is more likely that their magnetic fields are of fossil origin. The comparison gives further support for a dynamo origin of the magnetic field and activity of 37 Com.

\begin{acknowledgements}
We thank the TBL and CFHT teams for providing service observing with Narval and ESPaDOnS. This work is supported by the OP Human Resources Development, ESF and Republic of Bulgaria, project BG051PO001-3.3.06-0047. The observations in 2008 were funded under an OPTICON program. The observations in 2010 with Narval were funded under Bulgarian NSF grant DSAB 02/3/2010. R.K.-A. and S.Ts. acknowledge the mobility support under the Rila program DRILA 01/3. G.A.W. acknowledges support from the Natural Sciences and Engineering Research Council of Canada (NSERC). C.C. acknowledges support from the Swiss National Foundation (FNS) and the French Programme National de Physique Stellaire (PNPS) of CNRS/INSU. N.A.D. thanks Saint Petersburg State University, Russia, for research grant 6.38.18.2014 and FAPERJ, Rio de Janeiro - Brazil, for Visiting Researcher Grant E-26/200.128/2015.
\end{acknowledgements}


\begin{thebibliography}{}

\bibitem[2013]{Adamczak} Adamczak, J. \& Lambert, D.L. 2013, ApJ, 765, 155

\bibitem[1989]{Anders1989} Anders, E. \& Grevesse, N. 1989, Geochim. Cosmochim. Acta, 53, 197

\bibitem[2005]{Asplund2005} Asplund, M., Grevesse, N. \& Sauval, A.J. 2005, ASPC, 336, 25

\bibitem[2003]{Auriere2003} Auri\`ere, M. 2003, in ``Magnetism and Activity of the Sun and Stars'', Eds J. Arnaud and N. Meunier, EAS Publ. Series 9, 105

\bibitem[2007]{Auriere2007} Auri\`ere, M., Wade, G.A., Silvester, J. et al. 2007, A\&A, 475, 1053

\bibitem[2009]{Auriere2009a} Auri\`ere, M., Konstantinova-Antova, R., Petit, P., Wade, G. \& Roudier, T. 2009~a, IAUS, 259, 431

\bibitem[2009]{Auriere2009b} Auri\`ere, M., Wade, G.A., Konstantinova-Antova, R. et al. 2009~b, A\&A, 504, 231

\bibitem[2011]{Auriere2011} Auri\`ere, M., Konstantinova-Antova, R., Petit, P. et al. 2011, A\&A, 534, 139

\bibitem[2012]{Auriere2012} Auri\`ere, M., Konstantinova-Antova, R., Petit, P. et al. 2012, A\&A, 543, 118

\bibitem[2015]{Auriere2015} Auri\`ere, M., Konstantinova-Antova, R., Charbonnel, C. et al. 2015, A\&A, 574, 90

\bibitem[2015]{Braithwaite} Braithwaite, J. \& Spruit, H.C. 2015, arXiv:1510.03198

\bibitem[2016]{Borisova} Borisova, A., Auri\`ere, M., Petit, P. et al. 2016, A\&A, 591, 57

\bibitem[1989]{Brown89} Brown, J.A., Sneden, C., Lambert, D.L. \& Dutchover, E., Jr. 1989, ApJS, 71, 293

\bibitem[1991]{Brown91} Brown, S.F., Donati, J.-F., Rees, D.E. \& Semel, M. 1991, A\&A, 250, 463

\bibitem[2010]{Charbonnel2010} Charbonnel, C. \& Lagarde, N. 2010, A\&A, 522, 10


\bibitem[2011]{Claret2011} Claret, A. \& Bloemen, S. 2011, A\&A, 529A, 75C

\bibitem[1990]{Costes1990} Costes, M., Naulin, C. \& Dorthe, G. 1990, A\&A, 232, 270

\bibitem[2012]{deBruijne} de Bruijne, J.H.J. \& Eilers, A.-C. 2012, A\&A, 546, 61

\bibitem[2009]{Decressin09} Decressin, T., Mathis, S., Palacios, A. et al. 2009, A\&A, 495, 271

\bibitem[1999]{deMedeirosMayor} de Medeiros, J.R. \& Mayor, M. 1999, A\&AS, 139, 433

\bibitem[1999]{deMedeiros} de Medeiros, J.R., Konstantinova-Antova, R.K. \& da Silva, J.R.P. 1999, A\&A, 347, 550

\bibitem[1997]{Donati1997a} Donati, J.-F. \& Brown, S. F. 1997, A\&A, 326, 1135

\bibitem[1997]{Donati1997b} Donati, J.-F., Semel, M., Carter, B.D., Rees, D.E. \& Collier Cameron, A. 1997, MNRAS, 291, 658

\bibitem[2003]{Donati2003} Donati, J.-F., Collier Cameron, A., Semel, M. et al. 2003, MNRAS, 345, 1145

\bibitem[2006]{Donati2006a} Donati, J.-F., Catala C., Landstreet J. \& Petit P. 2006~a, in Casini R., Lites B., eds, Solar Polarization Workshop n4 Vol.358 of ASPC series, 362

\bibitem[2006]{Donati2006b} Donati, J.-F., Howarth, I.D., Jardine, M.M. et al. 2006~b, MNRAS, 370, 629

\bibitem[1994]{Drake1994} Drake, J.J. \& Lambert, D.L. 1994, ApJ, 435, 797

\bibitem[2008]{Drake2008} Drake, N.A. \& Pereira, C.B. 2008, AJ, 135, 1070

\bibitem[1991]{Duncan} Duncan, D.K., Vaughan, A.H., Wilson, O.C. et al. 1991, ApJS, 76, 383

\bibitem[1997]{Fekel} Fekel, F.C. 1997, PASP, 109, 514

\bibitem[1996]{Flower} Flower, P.J. 1996, ApJ, 469, 355

\bibitem[2016]{Folsom} Folsom, C.P., Petit, P., Bouvier, J. et al. 2016, MNRAS, 457, 580



\bibitem[1999]{Hobbs1999} Hobbs, L.M., Thorburn, J.A. \& Rebull, L.M. 1999, ApJ, 523, 797

\bibitem[1992]{Huang1992} Huang, Y., Barts, S.A. \& Halpern, J.B. 1992, J. Phys. Chem., 96, 425

\bibitem[1989]{Keenan} Keenan, P.C. \& McNeil, R.C. 1989, ApJS, 71, 245

\bibitem[2010]{Kochukhov} Kochukhov, O. \& Wade, G.A. 2010, A\&A, 513, 13

\bibitem[2008a]{Konst2008a} Konstantinova-Antova, R., Auri\`ere, M., Alecian, E. et al. 2008~a, AIPC, 1043, 405

\bibitem[2008b]{Konst2008b} Konstantinova-Antova, R., Auri\`ere, M., Iliev, I.K. et al. 2008~b, A\&A, 480, 475

\bibitem[2009]{Konst2009} Konstantinova-Antova, R., Auri\`ere, M., Schr\"{o}der, K.-P. \& Petit, P. 2009, IAUS, 259, 433

\bibitem[2012]{Konst2012} Konstantinova-Antova, R., Auri\`ere, M., Petit, P. et al. 2012, A\&A, 541, 44

\bibitem[2014]{Konst2014} Konstantinova-Antova, R., Auri\`ere, M., Charbonnel, C. et al. 2014, Proceedings IAU Symposium No. 302

\bibitem[1999]{Kupka1999} Kupka, F., Piskunov, N., Ryabchikova, T.A., Stempels, H.C., Weiss, W.W. 1999, A\&AS, 138, 119

\bibitem[1993]{Kurucz13} Kurucz, R.L. 1993, CD-ROM 13, Atlas9 Stellar Atmosphere Programs and 2~km\,s$^{-1}$ Grid (Cambridge: Smithsonian Astrophys. Obs.)

\bibitem[1993]{Kurucz18} Kurucz, R.L. 1993, SAO, Cambridge, CDROM 18


\bibitem[1981]{Lambert1981} Lambert, D.L. \& Ries, L.M. 1981, ApJ, 248, 228

\bibitem[2014]{Marsden} Marsden, S.C., Petit, P., Jeffers, S.V. et al. 2014, MNRAS, 444, 3517

\bibitem[1990]{McWilliam} McWilliam, A. 1990, ApJS, 74, 1075

\bibitem[2008]{Morin08} Morin, J., Donati, J.-F., Petit, P. et al. 2008, MNRAS, 390, 567

\bibitem[2010]{Morin10} Morin, J., Donati, J.-F., Petit, P. et al. 2010, MNRAS, 407, 2269

\bibitem[2012]{Morgenthaler} Morgenthaler, A., Petit, P., Saar, S. et al. 2012, A\&A, 540, 138

\bibitem[2007]{Moutou} Moutou, C., Donati, J.-F., Savalle, R. et al. 2007, A\&A, 473, 651

\bibitem[2012]{Palacios1} Palacios, A., Parthasarathy, M., Bharat Kumar, Y. \& Jasniewicz, G. 2012, A\&A, 538, 68

\bibitem[2016]{Palacios2} Palacios, A., Jasniewicz, G., Masseron, T., Th\'evenin, F., Pasquet, J., Parthasarathy, M. 2016, A\&A, 587, A42

\bibitem[1997]{Perryman} Perryman, M.A.C., Lindegren, L., Kovalevsky, J. et al. 1997, A\&A, 323, 49

\bibitem[2002]{Petit2002} Petit, P., Donati, J.-F. \& Collier Cameron, A. 2002, MNRAS, 334, 374

\bibitem[2004]{Petit2004} Petit, P., Donati, J.-F., Oliveira, J.M. et al. 2004, MNRAS, 351, 826


\bibitem[1994]{Pradhan1994} Pradhan, A.D., Partridge, H. \& Bauschlicher, C.W. 1994, J. Chem. Phys., 101, 3857

\bibitem[1979]{Rees} Rees, D.E. \& Semel, M.D. 1979, A\&A, 74, 1

\bibitem[1975]{Reimers1975} Reimers, D. 1975, Memoires of the Societe Royale des Sciences de Liege, 8, 369

\bibitem[2015]{See} See, V., Jardine, M., Vidotto, A.A. et al. 2015, MNRAS, 453, 4301

\bibitem[1989]{Semel1989} Semel, M. 1989, A\&A, 225, 456

\bibitem[1993]{Semel1993} Semel, M., Donati, J.-F. \& Rees, D.E. 1993, A\&A, 278, 231

\bibitem[2006]{Siess06} Siess, L. 2006, A\&A, 448, 717

\bibitem[2012]{Silvester} Silvester, J., Wade, G.A., Kochukhov, O. et al. 2012, MNRAS, 426, 1003

\bibitem[1984]{Skilling} Skilling, J. \& Bryan, R.K. 1984, MNRAS, 211, 111

\bibitem[1998]{Smith1998} Smith, V.V., Lambert, D.L. \& Nissen, P.E. 1998, ApJ, 506, 405

\bibitem[1973]{Sneden1973} Sneden, C. 1973, Ph.D. Thesis, Univ. of Texas

\bibitem[1978]{Sneden1978} Sneden, C., Lambert, D.L., Tomkin, J. \& Peterson, R.C. 1978, ApJ, 222, 585S


\bibitem[2010]{Soubiran} Soubiran, C., Le Campion, J.-F., Cayrel de Strobel, G. \& Caillo, A. 2010, A\&A, 515, 111

\bibitem[1988]{Strassmeier} Strassmeier, K.G. \& Hall, D.S. 1988, ApJS, 67, 439

\bibitem[2011]{Strugarek} Strugarek, A., Brun, A.S. \& Zahn, J.-P. 2011, A\&A, 532, 34

\bibitem[2008]{Tokovinin} Tokovinin, A. 2008, MNRAS, 389, 925


\bibitem[2013]{Tsvetkova} Tsvetkova, S., Petit, P., Auri\`ere, M. et al. 2013, A\&A, 556, 43

\bibitem[2007]{vanLeeuwen} van Leeuwen, F. 2007, A\&A, 474, 653

\bibitem[2000]{Wade2000a} Wade, G.A., Donati, J.-F., Landstreet, J.D. \& Shorlin, S.L.S. 2000~a, MNRAS, 313, 823

\bibitem[2000]{Wade2000b} Wade, G.A., Donati, J.-F., Landstreet, J.D. \& Shorlin, S.L.S. 2000~b, MNRAS, 313, 851


\end{thebibliography}
\end{document}